\documentclass{JINST}
\usepackage{epsfig}
\usepackage{graphicx}
\usepackage{amsbsy}
\usepackage{amssymb,fontenc,times,mathptmx}
\usepackage{rotating}
\title{
\centering
Study of the interactions of pions in the CALICE silicon-tungsten calorimeter prototype
}
\author{\centering 
\LARGE\bf The CALICE Collaboration
}
\author{\centering
C.\,Adloff, 
Y.\,Karyotakis
\\ \it
Laboratoire d'Annecy-le-Vieux de Physique des Particules, Universit\'{e} de Savoie,
CNRS/IN2P3,
9~Chemin de Bellevue BP110, F-74941 Annecy-le-Vieux Cedex, France
}

\author{\centering
J.\,Repond
\\ \it
Argonne National Laboratory,
9700 S.\ Cass Avenue,
Argonne, IL 60439-4815,
USA
}

\author{\centering
J.\,Yu
\\ \it
Department of Physics, SH108, University of Texas, Arlington, TX 76019, USA
}

\author{\centering
G.\,Eigen
\\ \it
University of Bergen, Inst.\, of Physics, Allegaten 55, N-5007 Bergen, Norway
}

\author{\centering
Y.\,Mikami, 
N.\,K.\,Watson, 
J.\,A.\,Wilson
\\ \it
University of Birmingham,
School of Physics and Astronomy,
Edgbaston, Birmingham B15 2TT, UK
}

\author{\centering 
T.\,Goto, 
G.\,Mavromanolakis\footnote{Now at CERN}, 
M.\,A.\,Thomson, 
D.\,R.\,Ward, 
W.\,Yan\footnote{Now at Dept.\, of Modern Physics, Univ. of Science and Technology of China, 96 Jinzhai Road, Hefei, Anhui, 230026, P.\, R.\, China}
\\ \it
University of Cambridge, Cavendish Laboratory, J J Thomson Avenue, CB3 0HE, UK
}

\author{\centering 
D.\,Benchekroun, 
A.\,Hoummada, 
Y.\,Khoulaki
\\ \it
Universit\'{e} Hassan II A\"{\i}n Chock, Facult\'{e} des sciences.\, B.P. 5366 Maarif, Casablanca, Morocco
}

\author{\centering 
J.\,Apostolakis, 
A.\,Ribon,
V.\,Uzhinskiy
\\ \it 
CERN, 1211 Gen\`{e}ve 23, Switzerland
}

\author{\centering
M.\,Benyamna, 
C.\,C\^{a}rloganu,
F.\,Fehr,  
P.\,Gay
\\ \it
Laboratoire de Physique Corpusculaire de Clermont-Ferrand (LPC),
Universit\'e Blaise Pascal, CNRS/IN2P3, 
24 avenue des Landais,
63177 Aubi\`ere CEDEX, France
}

\author{\centering
G.\,C.\,Blazey,
D.\,Chakraborty,
A.\,Dyshkant,
K.\,Francis,
D.\,Hedin,
J.\,G.\,Lima,
V.\,Zutshi
\\ \it
NICADD, Northern  Illinois University,
Department of Physics,
DeKalb, IL 60115,
USA
}

\author{\centering 
J.\,-Y.\,Hostachy, 
K.\,Krastev, 
L.\,Morin
\\ \it
Laboratoire de Physique Subatomique et de Cosmologie - Universit\'{e} Joseph Fourier Grenoble 1 -
CNRS/IN2P3 - Institut Polytechnique de Grenoble,
53, rue des Martyrs,
38026 Grenoble CEDEX, France
}

\author{\centering 
N.\,D'Ascenzo, 
U.\,Cornett, 
D.\,David, 
R.\,Fabbri, 
G.\,Falley, 
K.\,Gadow, 
E.\,Garutti,
P.\,G\"{o}ttlicher, 
T.\,Jung, 
S.\,Karstensen, 
A.\,-I.\,Lucaci-Timoce,
B.\,Lutz, 
N.\,Meyer,
V.\,Morgunov, 
M.\,Reinecke, 
F.\,Sefkow, 
P.\,Smirnov,
A.\,Vargas-Trevino, 
N.\,Wattimena, 
O.\,Wendt
\\ \it
DESY, Notkestrasse 85,
D-22603 Hamburg, Germany
}

\author{\centering  
N.\,Feege, 
M.\,Groll, 
J.\,Haller, 
R.\,-D.\,Heuer, 
S.\,Morozov, 
S.\,Richter, 
J.\,Samson
\\ \it
Univ. Hamburg,
Physics Department,
Institut f\"ur Experimentalphysik,
Luruper Chaussee 149,
22761 Hamburg, Germany
}

\author{\centering 
A.\,Kaplan, 
H.\,-Ch.\,Schultz-Coulon, 
W.\,Shen, 
A.\,Tadday
\\ \it
 University of Heidelberg, Fakultat fur Physik und Astronomie,
Albert Uberle Str. 3-5 2.OG Ost,
D-69120 Heidelberg, Germany
}

\author{\centering 
B.\,Bilki, 
E.\,Norbeck, 
Y.\,Onel
\\ \it
University of Iowa, Dept. of Physics and Astronomy,
203 Van Allen Hall, Iowa City, IA~52242-1479, USA
}

\author{\centering 
E.\,J.\,Kim
\\ \it
Chonbuk National University, Jeonju, 561-756, South Korea
}

\author{\centering 
G.\,Kim, D-W.\,Kim, K.\,Lee, S.\,C.\,Lee
\\ \it
Kangnung National University, HEP/PD, Kangnung, South Korea
}

\author{\centering 
K.\,Kawagoe, Y.\,Tamura 
\\ \it
 Department of Physics, Kobe University, Kobe, 657-8501, Japan
}

\author{\centering 
P.\,D.\,Dauncey, 
A.\,-M.\,Magnan, 
H.\,Yilmaz, 
O.\,Zorba
\\ \it
Imperial College, Blackett Laboratory,
Department of Physics,
Prince Consort Road,
London SW7 2AZ, UK 
}

\author{\centering 
V.\,Bartsch\footnote{Now at University of Sussex, Physics and Astronomy Department, Brighton, Sussex, BN1 9QH, UK}, 
M.\,Postranecky, M.\,Warren, M.\,Wing
\\ \it
Department of Physics and Astronomy, University College London,
Gower Street,
London~WC1E~6BT, UK
}

\author{\centering 
M.\,G.\,Green, 
F.\,Salvatore\footnotemark[3]
\\ \it
Royal Holloway University of London,
Dept. of Physics,
Egham, Surrey TW20 0EX, UK
}

\author{\centering 
M.\,Bedjidian,   
R.\,Kieffer, 
I.\,Laktineh
\\ \it
Universit\'{e} de Lyon, F-69622, Lyon, France ;
Universit\'{e} de Lyon 1, Villeurbanne ;
CNRS/IN2P3, Institut de Physique Nucl\'{e}aire de Lyon
}

\author{\centering 
M.-C.\,Fouz
\\ \it
CIEMAT, Centro de Investigaciones Energeticas, Medioambientales y Tecnologicas, Madrid. Spain 
}

\author{\centering 
D.\,S.\,Bailey, 
R.\,J.\,Barlow, 
M.\,Kelly, 
R.\,J.\,Thompson 
\\ \it
The University of Manchester, School of Physics and Astronomy,
Schuster Lab,
Manchester~M13~9PL,
UK
}

\author{\centering 
M.\,Danilov, 
E.\,Tarkovsky
\\ \it
Institute of Theoretical and Experimental Physics, B. Cheremushkinskaya ul. 25,
RU-117218 Moscow, Russia
}

\author{\centering 
N.\,Baranova,
D.\,Karmanov, 
M.\,Korolev, 
M.\,Merkin,
A.\,Voronin
\\ \it
M.V.Lomonosov Moscow State University, D.V.Skobeltsyn Institute of Nuclear
Physics (SINP~MSU),
1/2 Leninskiye Gory, Moscow, 119991, Russia
}

\author{\centering 
A.\,Frey\footnote{Now at Univ.\, of G\"{o}ttingen}, 
S.\,Lu\footnote{Now at DESY}, 
K.\,Seidel, 
F.\,Simon,
C.\,Soldner, 
L.\,Weuste
\\ \it
Max Planck Inst. f\"ur Physik,
F\"ohringer Ring 6,
D-80805 Munich, Germany
}

\author{\centering 
J.\,Bonis, 
B.\,Bouquet,    
S.\,Callier, 
P.\,Cornebise, 
Ph.\,Doublet,
M.\,Faucci Giannelli, 
J.\,Fleury,
H.\,Li,  
G.\,Martin-Chassard, 
F.\,Richard, 
Ch.\,de la Taille, 
R.\,Poeschl, 
L.\,Raux,  
N.\,Seguin-Moreau, 
F.\,Wicek
\\ \it
Laboratoire de L'acc\'elerateur Lin\'eaire,
Centre d'Orsay, Universit\'e de Paris-Sud XI,
BP 34, B\^atiment 200,
F-91898 Orsay CEDEX, France
}

\author{\centering 
M.\,Anduze, 
V.\,Boudry, 
J-C.\,Brient, 
G.\,Gaycken,
D.\,Jeans, 
P.\,Mora de Freitas, 
G.\,Musat, 
M.\,Reinhard, 
A.\,Roug\'{e},
M.\,Ruan,  
J-Ch.\,Vanel, 
H.\,Videau
\\ \it
      Laboratoire Leprince-Ringuet (LLR)  -- \'{E}cole Polytechnique,
      CNRS/IN2P3,
      Palaiseau, F-91128 France
}

\author{\centering 
K-H.\,Park
\\ \it
Pohang Accelerator Laboratory, Pohang 790-784, South Korea
}

\author{\centering 
J.\,Zacek 
\\ \it
Charles University, Institute of Particle \& Nuclear Physics,
V Holesovickach 2,
CZ-18000 Prague~8, Czech Republic  
}

\author{\centering 
J.\,Cvach, 
P.\,Gallus, 
M.\,Havranek, 
M.\,Janata, 
M.\,Marcisovsky, 
I.\,Polak, 
J.\,Popule, 
L.\,Tomasek, 
M.\,Tomasek, 
P.\,Ruzicka, 
P.\,Sicho, 
J.\,Smolik, 
V.\,Vrba, 
J.\,Zalesak 
\\ \it
Institute of Physics, Academy of Sciences of the Czech Republic, Na Slovance 2,
CZ-18221 Prague 8, Czech Republic
}

\author{\centering 
B.\,Belhorma, M.\,Belmir
\\ \it
Centre National de l'Energie, des Sciences et des Techniques Nucl\'{e}aires, 
B.P. 1382, R.P. 10001, Rabat, Morocco
}

\author{\centering 
S.\,W.\,Nam, I.\,H.\,Park, J.\,Yang 
\\ \it
Ewha Womans University, Dept. of Physics,
Seoul 120,
South Korea
}

\author{\centering 
Jong-Seo Chai, Jong-Tae Kim, Geun-Bum Kim
\\ \it
Sungkyunkwan University,
300 Cheoncheon-dong, Jangan-gu, Suwon, Gyeonggi-do  440-746, South Korea
}

\author{\centering 
J.\,Kang, Y.\,-J.\,Kwon  
\\ \it
Yonsei  University, Dept. of Physics,
134 Sinchon-dong,
Sudaemoon-gu, Seoul 120-749,
South~Korea
}

\abstract{ A prototype silicon-tungsten electromagnetic calorimeter for an ILC
detector was tested in 2007 at the CERN SPS test beam.  Data were collected with
electron and hadron beams in the energy range 8 to 80 GeV.  The analysis
described here focuses on the interactions of pions in the calorimeter.  One of
the main objectives of the CALICE program is to validate the Monte Carlo tools
available for the design of a full-sized detector.  The interactions of pions in
the Si-W calorimeter are therefore confronted with the predictions of various
physical models implemented in the {\tt GEANT4} simulation framework.  }


\begin{document}

\section{Introduction}
\label{sect:Intro}
The current designs of the calorimetry for detectors at the ILC are
in large part driven by the demands of jet energy reconstruction.   
A key aim is to achieve a relative energy resolution of 
$\sim30\%/\sqrt{E/\mathrm{GeV}}$; this precision would for example 
permit the reconstruction of the hadronic decays of
the W and Z bosons with a precision comparable with their natural widths,
and would thus allow the W and Z to be distinguished in their hadronic (two-jet) 
decay modes.  

This  target for jet energy resolution 
is roughly a factor two better than achieved in previous detectors.  
The most promising way to achieve it is believed to be through the
``{\it particle flow}''approach~\cite{ref:pflow}.  The
idea is that particles of different types in jets should be 
reconstructed in the different parts of the detector where they can 
be measured most precisely: the charged particles in the tracking system, 
photons in the electromagnetic calorimeter (ECAL) and neutral hadrons in 
a combination of the ECAL and the 
hadronic calorimeter (HCAL).  
The key to this concept is to minimise confusion in the pattern recognition 
in the calorimeters, and to achieve this, 
high spatial granularity is required.  
The optimum design of the ILC detectors can be addressed by Monte Carlo 
simulation, but in order to do this, it is crucial first to validate the Monte 
Carlo tools against data.  

A first round of beam tests was performed at DESY and CERN in summer 2006, 
followed by more complete tests in 2007.  These tests were performed with 
a combined system of a silicon-tungsten (Si-W) ECAL, 
followed by a hadronic calorimeter and then 
a coarser tail catcher, the latter two 
both built from a scintillator-iron sandwich 
structure.  Since the ECAL has a length $\sim1.0\lambda_{\mathrm{int.}}$, 
it will seldom fully contain a hadronic shower.  However, more than half 
of hadronic showers will start their development in the ECAL, 
so it is important to investigate the properties of hadronic showers
initiated  in the (mainly tungsten) material of the ECAL.  The ECAL 
also has some advantages for the study of hadronic shower properties.
For example, the interaction point can be identified with good precision.  
The small ratio of $X_0/\lambda_{\mathrm{int.}}$ (a factor $\sim3$ smaller 
for tungsten than for iron) means 
that the electromagnetic products of the first interaction can be 
expected to shower rapidly after the interaction point, and we can
hope to exploit the fine granularity of the detector to separate them
from the hadronic products.  This means that the longitudinal shower 
development after the interaction point is potentially 
a particularly interesting quantity.       

In Sect.~\ref{sect:ECAL} we briefly describe the main features of 
the Si-W calorimeter and in Sect.~\ref{sect:TB} we outline the test 
beam setup at CERN.  Then in Sect.~\ref{sect:MC} we review the 
Monte Carlo models with which the data are confronted.  In 
Sect.~\ref{sect:Sel} we explain the selection of data 
for comparison with simulation in Sect.~\ref{sect:Comp}.

\section{The CALICE Si-W ECAL prototype}
\label{sect:ECAL}
The ECAL prototype used in this study 
is a silicon-tungsten sampling calorimeter, 
made of 30 readout layers~\cite{ref:ECALcomm} . 
The active detectors were silicon wafers segmented into 
a 6$\times$6 array of diode pads, each 
with a size of 1$\times$1~cm$^2$. 
The mechanical structure was built from tungsten sheets 
wrapped in carbon fibre. 
Between each sheet of tungsten, the structure contained alveolar slots, 
into which detector slabs were inserted, each consisting of
a further tungsten sheet sandwiched between two layers of sensors.   
The prototype was constructed from three stacks, 
each composed of ten layers of alternating tungsten and silicon,  
and each stack having a different tungsten thickness: 
1.4 mm or $0.4X_0$ per layer in the first stack, 
2.8 mm or $0.8X_0$ per layer in the second stack and
4.2 mm or $1.2X_0$ per layer in the rear stack. 
In terms of interaction lengths,  these tungsten layers each contribute
$\sim1.4\%\lambda_{\mathrm{int.}}$,
$\sim2.7\%\lambda_{\mathrm{int.}}$ and $\sim4.1\%\lambda_{\mathrm{int.}}$ respectively.
Other materials (mainly the carbon-fibre--epoxy mechanical structure, PCBs and the silicon itself) 
contribute alternately $\sim0.2\%\lambda_{\mathrm{int.}}$ and $\sim1.0\%\lambda_{\mathrm{int.}}$ 
between successive samplings.  
The overall thickness is about 20 cm, corresponding to  
$\sim24.6\,X_0$ or $\sim1.0\lambda_{\mathrm{int.}}$ at normal incidence.

A detailed description of the prototype's hardware and of its commissioning 
in test beams at DESY and CERN in 2006, can be found in ref.~\cite{ref:ECALcomm}, 
to which the reader is referred for much fuller details. 

The full prototype consists of a 3$\times$3 array of wafers (i.e.\ an 18$\times$18 array of
1~cm$^2$ pads) in each layer.
The version of the prototype tested at CERN in 2007 consisted of 30 layers, 
of which initially the first twelve were instrumented 
with a 3$\times$2 array of wafers, and the remaining eighteen were fully 
equipped with a 3$\times$3 array.  Later in the 2007 run six more layers were completed, 
leaving only the first six in a 3$\times$2 configuration.  
  
In offline analysis, the raw hit energies in each cell have their 
pedestals subtracted, and are converted from raw ADC counts into 
Minimum Ionising Particle equivalents (MIPs), as explained in ref.~\cite{ref:ECALcomm}.  
This gain correction is derived by finding the most probable energy deposition in each cell 
in data recorded with a high energy muon beam~\cite{ref:ECALcomm}, with a typical relative 
precision of $\sim0.5\%$ per pad.  

\section{The test beam}
\label{sect:TB}

\begin{figure}[tbp]
\centering \includegraphics[width=0.9\textwidth]{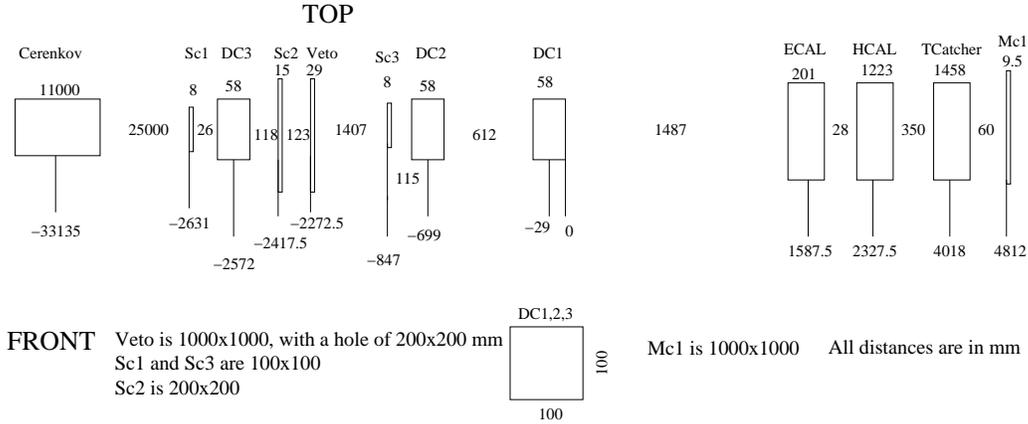}
\caption{\label{fig:TBlayout}\em Schematic layout for the CALICE beam tests at CERN in 2007 (not to scale).  
Dimensions are indicated in mm. }
\end{figure}

In this paper we report on some of the data taken in 2007 in the CERN H6 test beam~\cite{ref:H6}.
The layout of the CALICE calorimeters in the test beam is shown schematically in 
figure~\ref{fig:TBlayout}.  The three calorimeters used were the Si-W ECAL, 
followed by an HCAL prototype~\cite{ref:HCALcomm} (using iron as the absorber and 38 layers of 
scintillator tiles with analogue readout
as the active medium, using tile size 3$\times$3~cm in the shower core) 
and a tail catcher (TCMT - 
also an iron calorimeter with sixteen layers of 5~cm wide 
scintillator strips).  The thickness of the iron sheets was $\sim18$~mm in the HCAL,
 $\sim20$~mm for the first eight layers of the TCMT, and  
$\sim100$~mm in the rear section of the TCMT.
The ECAL and HCAL were mounted on a movable stage, providing the possibility to 
translate and rotate the calorimeters with respect to the beam. 

Upstream of the calorimeters were scintillation counters used for triggering.  
A muon veto counter was also installed downstream of the TCMT.   The standard trigger 
used for the present analysis required a coincidence between any
two of the upstream counters  Sc1, Sc2 and Sc3. In addition, three sets of wire chambers
were operated to permit the beam position and direction to be measured upstream of the calorimeters.
A \v{C}erenkov counter was used in threshold mode to assist particle identification.

Data were recorded in 2007 using electron, hadron and muon beams with 
energies in the range 8 to 180 GeV.  A variety of different calorimeter translations 
and rotations were used, and in all $\sim$200 million triggers were recorded, 
including calibration data.  For the present study, we used data corresponding to
a configuration with the beam impinging at normal incidence close 
to the centre of the calorimeters.  The analysis uses $\pi^-$ beams at energies of 
8, 10, 12, 15 and 20~GeV, and $\pi^+$ runs at 30, 50 and 80~GeV.  Each run 
comprised typically $\sim$200,000 triggers and $\sim$100,000 hadronic events.  
For the $\pi^-$ runs 
the pressure in the \v{C}erenkov counter was set so that electrons could be vetoed, 
while for the $\pi^+$ runs, it  was set so as to separate protons from pions. 

\section{Monte Carlo simulation}
\label{sect:MC}
The main purpose of the present analysis is to confront the data with Monte Carlo 
simulations, and specifically to test the different physical models available.  
These simulations were carried out in the framework of {\tt GEANT4}~\cite{ref:G4},
using version 4.9.3, released in December 2009.  The results presented are based on 
samples of typically 50000 simulated events.

In order to describe the geometry of the CALICE prototypes within {\tt GEANT4} we used the 
program Mokka~\cite{ref:mokka}, 
which is also capable of simulating full ILC detector geometries.
For the current study, the simulation of the ECAL is the most important.  Individual silicon 
sensors, with their pad structure and guard rings, are represented in Mokka, 
as well as the tungsten 
radiator, and other passive materials such as the carbon-fibre and epoxy of the 
support structure and the PCBs used for readout. The layer-by-layer staggering of the 
wafer positions is simulated.  
Similar representations of the HCAL and TCMT are also included.
The detectors upstream of the calorimeter systems (\v{C}erenkov, scintillators 
and tracking chambers) are also modelled by Mokka.  

During particle transport in {\tt GEANT4}, Mokka records the ionisation energy deposited in 
sensitive detectors (Silicon pads, scintillators etc.).  
A simple simulation of the noise contribution in each cell is performed, 
as outlined in ref.~\cite{ref:ECALresp}. 
At the end of the simulation, 
the summed energy deposits in each cell are finally converted into MIPs.   
The conversion
factor is based on matching the minimum ionising energy peak in data and simulation.
This procedure is estimated to lead to a $\sim 1\%$ systematic uncertainty in the 
relative energy scales of the data and simulation. 
As a first step in the analysis, 
a hit energy cut of 0.6 MIPs is imposed in both simulation and data, 
in order to eliminate most noise-only hits.  The hits are also required to occur within 
150~ns of the beam arrival time at the trigger counters, in order to emulate the
behaviour of the readout electronics. 

\begin{figure}[tbp]
\centering \includegraphics[width=0.78\textwidth, angle=0]{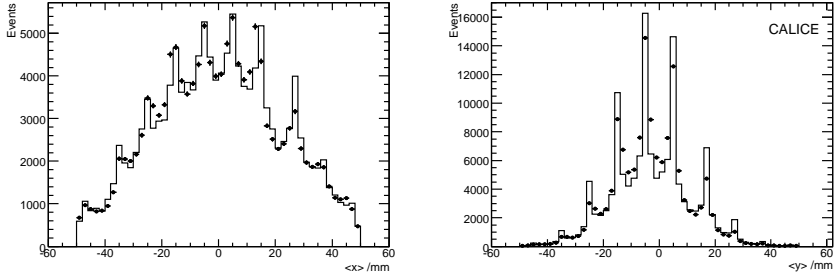}
\caption{\label{fig:BeamProfiles}\em Example of the beam profiles in $x$ and $y$
(based on the shower centroid reconstructed in the ECAL) 
observed with a 20~GeV $\pi^-$ beam.  Data (points with error bars)
 are compared with the tuned simulation (solid histogram). 
The distributions are normalised to the same numbers of events. }
\end{figure}

In the simulation, pions are simulated starting with a Gaussian transverse profile in $x$ and 
$y$,\footnote{The CALICE coordinate system is defined with $z$ along the nominal beam direction, 
$y$ vertical, and $x$ horizontal.  The origin is at the centre of the exit face of the
most downstream tracking chamber.}  
at a point $\sim$60~m upstream of the calorimeters, so that they pass through, and potentially 
interact in, the \v{C}erenkov counter, 
trigger scintillators, tracking chambers and intervening air.  
Hits are recorded in the scintillators and 
tracking chambers, and energies deposited in the scintillators are used as the basis of a simple 
simulation of the trigger.  The beam profile in the ECAL, defined by the distributions of the
shower barycentre in $x$ and $y$, is then compared between data and simulation.  The input 
Gaussian parameters are adjusted until satisfactory agreement is achieved.  A typical 
example of the comparison after tuning is shown in figure~\ref{fig:BeamProfiles}.

{\tt GEANT4} provides the user with a number of ``physics lists'' --- combinations of models 
selected to simulate particle interactions for different species at different energies. 
In our study we are particularly concerned with the hadronic interactions for pions. 
Typically these physics lists use different models in different energy ranges, 
with smooth interpolation between models achieved by allowing the energy ranges to overlap, 
with smoothly varying random selection between the models in the changeover region.  
Detailed descriptions of the models can be found in ref.~\cite{ref:G4PhysRef}.
The following physics lists have been chosen for our investigation, based on 
advice and recommendations from the {\tt GEANT4} authors:
\begin{description}
\item[\tt LHEP]Uses the {\tt LEP} (low energy parametrised) and {\tt HEP} (high energy parametrised) 
models, making a transition between the two over the range 25--55~GeV.  
These are essentially a recoding into {\tt C++} 
of the {\tt GHEISHA} model~\cite{ref:Gheisha} extensively used in earlier simulations 
using {\tt GEANT3}, for example.  It is still widely used, even though it is not regarded as 
the state-of-the-art choice.  
\item[\tt FTFP\_BERT]
Uses the {\tt GEANT4} implementation of the Bertini cascade model\footnote{
This model incorporates the Bertini intra-nuclear cascade model with excitons, 
a pre-equilibrium model,
a nucleus explosion model, a fission model, and an evaporation model.}~\cite{ref:Bertini} 
for low energies, 
making a transition to the {\tt FTFP} model, 
based on the {\tt GEANT4} implementation of the Fritiof diffractive  string model~\cite{ref:Fritiof}, 
at pion energies between 4 and 5~GeV.  We have also studied the {\tt FTFP\_BERT\_TRV} list, which has 
the same model content as {\tt FTFP\_BERT}, but with a higher transition energy, 
in the region  6--8~GeV.  In practice, for the data presented here, 
we find no significant differences between these two physics lists, so we only show {\tt FTFP\_BERT} here.
\item[\tt QGSP\_BERT]
Uses the Bertini model~\cite{ref:Bertini} at low energies, 
making a transition to the {\tt LEP} ({\tt GHEISHA}) 
model between 9.5 and 9.9~GeV, and a further transition to the {\tt QGSP} model 
between 12 and 25~GeV.  {\tt QGSP} uses a {\tt GEANT4} implementation of a 
string model~\cite{ref:QGS} for the high energy interaction,
supplemented by the {\tt GEANT4} precompound model\cite{ref:PreCompound} 
describing deexcitation of the nucleus.
We have also studied the {\tt QGSP\_BERT\_HP} list, which has 
the same model content as {\tt QGSP\_BERT}, combined with a more accurate treatment of 
low energy neutron scattering.  For the distributions presented in this paper we 
find no significant benefit in using the {\tt QGSP\_BERT\_HP} list, and hence we only present 
{\tt QGSP\_BERT}.
\item[\tt QGSP\_BERT\_TRV] has the same model content as {\tt QGSP\_BERT}, 
but with the transition between {\tt LHEP} and {\tt QGSP} occurring at lower energy, 
in the region 10--15~GeV, in order to reduce the reliance on {\tt LHEP}. 
\item[\tt QGSP\_FTFP\_BERT] Similar to {\tt QGSP\_BERT}, with different transition energies, 
and using {\tt FTFP} instead of {\tt LEP} in the intermediate region.
\item[\tt QGS\_BIC]
Uses the {\tt GEANT4} binary cascade model ({\tt BIC})~\cite{ref:BIC} at the lowest energies, 
then {\tt LEP} in the intermediate region, and {\tt QGS} at high energies. The {\tt BIC} model
is also used for the rescattering of secondaries in this case (denoted {\tt QGSB} below) and the
{\tt GEANT4} precompound model is used to describe deexcitation of the nucleus.
\item[\tt QGSP\_BIC]
In this case the {\tt BIC} model is not used for pions (only for neutrons and protons), and
there is no rescattering of secondaries. {\tt LEP} is used in the low energy region, 
and {\tt QGSP} at high energies.
\item[\tt FTF\_BIC]
Uses the {\tt BIC} model at low energies, with a transition to  Fritiof between 4 and 5~GeV. 
The {\tt BIC} model is again used for the rescattering of secondaries in this case (denoted {\tt FTFB} below) and the
{\tt GEANT4} precompound model is used to describe deexcitation of the nucleus.
\end{description}
The physics content of these models for pions is summarised in table~\ref{tab:PhysListContent}. 
The {\tt QGSP\_BERT} physics list is generally favoured by the LHC general 
purpose detectors for calorimetry 
applications~\cite{ref:LHCsim}, based mainly on data from test beams of higher energies than those relevant for CALICE.  
In our analysis, we expect to be particularly sensitive 
to the model(s) used for the primary interaction 
at the energy of the incident beam.  
This is indicated for each physics list in table~\ref{tab:PhysListContent2}.

\begin{table}[hbp]\small
\centering
\caption{\label{tab:PhysListContent}\em The table shows the physics models invoked for pion inelastic interactions
in each physics list.  Where ranges overlap, {\tt GEANT4} chooses randomly between models, with probabilities varying linearly with energy over the range of overlap.  }
\begin{tabular}{|l||l|}
\hline
Physics List &  Model content (for $\pi^{\pm})$\\\hline
\tt FTFP\_BERT & Bertini (0--5~GeV); \tt FTFP \rm ($>$4~GeV) \\
\tt QGSP\_BERT & Bertini (0--9.9~GeV);  \tt LEP \rm (9.5--25~GeV); \tt QGSP \rm ($>$12~GeV) \\
\tt QGSP\_BERT\_TRV & Bertini (0--9.9~GeV);  \tt LEP \rm (9.5--15~GeV); \tt QGSP \rm ($>$10~GeV) \\
\tt QGSP\_FTFP\_BERT & Bertini (0--8~GeV); \tt FTFP \rm (6--25~GeV); \tt QGSP \rm ($>$12~GeV)  \\
\tt QGS\_BIC & \tt BIC \rm (0--1.3~GeV); \tt LEP \rm (1.2--25~GeV); \tt QGSB \rm ($>$12~GeV)\\
\tt QGSP\_BIC &  \tt LEP \rm ($<$25~GeV); \tt QGSP \rm ($>$12~GeV)\\
\tt FTF\_BIC & \tt BIC \rm (0--5~GeV); \tt FTFB \rm ($>$4~GeV)\\
\tt LHEP & \tt LEP \rm (0--55~GeV); \tt HEP \rm ($>$25~GeV) \\
\hline
\end{tabular}
\normalsize
\end{table}

\begin{table}[hbp]\small
\centering
\caption{\label{tab:PhysListContent2}\em 
Indicates the physics model(s) which will be  
used for the primary pion interaction at each of the beam energies considered here.
Where two models are indicated with a solidus, an appropriate random choice is made by 
{\tt GEANT4}. }
\scalebox{0.8}{
\begin{tabular}{|l||c|c|c|c|c|c|c|c|}
\hline
Physics List & 8 GeV & 10 GeV & 12 GeV & 15 GeV & 20 GeV & 30 GeV & 50 GeV & 80 GeV\\\hline\tt
\tt FTFP\_BERT & \tt FTFP & \tt FTFP & \tt FTFP & \tt FTFP & \tt FTFP & \tt FTFP & \tt FTFP & \tt FTFP  \\
\tt QGSP\_BERT & \tt BERT & \tt LEP & \tt LEP & \tt LEP/QGSP & \tt LEP/QGSP & \tt QGSP & \tt QGSP & \tt QGSP \\
\tt QGSP\_BERT\_TRV & \tt BERT & \tt LEP & \tt LEP/QGSP & \tt QGSP & \tt QGSP & \tt QGSP & \tt QGSP & \tt QGSP \\
\tt QGSP\_FTFP\_BERT & \tt FTFP & \tt FTFP & \tt FTFP & \tt FTFP/QGSP & \tt FTFP/QGSP &  \tt QGSP & \tt QGSP & \tt QGSP \\
\tt QGS\_BIC & \tt LEP & \tt LEP & \tt LEP & \tt LEP/QGSB & \tt LEP/QGSB & \tt QGSB  & \tt QGSB  & \tt QGSB \\
\tt QGSP\_BIC & \tt LEP & \tt LEP & \tt LEP & \tt LEP/QGSP & \tt LEP/QGSP & \tt QGSP  & \tt QGSP  & \tt QGSP \\
\tt FTF\_BIC & \tt FTFB & \tt FTFB & \tt FTFB & \tt FTFB & \tt FTFB & \tt FTFB & \tt FTFB & \tt FTFB \\
\tt LHEP & \tt LEP & \tt LEP & \tt LEP & \tt LEP & \tt LEP & \tt LEP/HEP & \tt LEP/HEP & \tt HEP  \\
\hline\hline
\end{tabular}
}
\normalsize
\end{table}

{\tt GEANT4} also provides several interesting physics lists based wholly or in part 
on the {\tt CHIPS} model~\cite{ref:CHIPS}.  
In this picture, the result of a hadronic or nuclear interaction is the
creation of a quasmon (essentially an intermediate state of excited
hadronic matter) which can
dissipate energy by radiating particles statistically
or by quark exchange with surrounding nucleons or clusters of nucleons.
We have made studies of several {\tt CHIPS}-based physics lists, but
we choose not to present comparisons at this stage because, as discussed in
Sect.~\ref{sect:Chips}, the model is still under active development.

\section{Selection of hadronic showers}
\label{sect:Sel}
\begin{figure}[tbp]
\centering \includegraphics[width=0.68\textwidth]{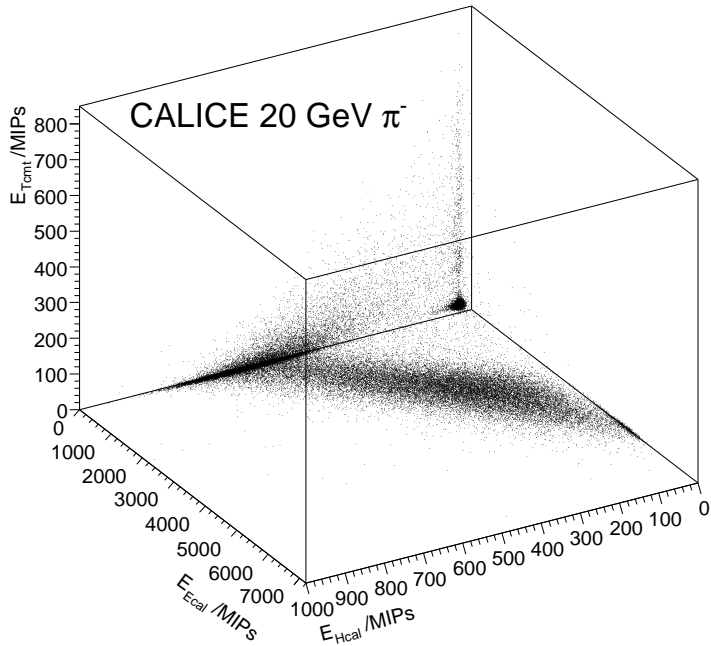}
\caption{\label{fig:MuonRemoval}\em Showing the principle of the cuts used to remove muons and 
non-interacting hadrons.  The energies in the three calorimeters (in MIPs) are plotted
for a 20~GeV data run.  The three stacks in the ECAL are combined using weights proportional to the 
corresponding thicknesses of tungsten, i.e.\ 1:2:3.}
\end{figure}

The principal task in the selection of events is to remove muons, 
electrons and protons from the sample.  Muons are characterised by 
a small energy deposition in all three calorimeters.  We show in 
figure~\ref{fig:MuonRemoval} a scatter plot of the energies seen in the 
three calorimeters.  A clear cluster close to the origin can be 
ascribed to muons.  Bands are also seen corresponding to sharing of 
energy between ECAL and HCAL, and between HCAL and TCMT, 
with a few showers only starting in the TCMT.   Accordingly, with a 10~GeV beam, 
we reject events for which the ECAL energy is less than 300~MIPs, the
HCAL energy is less than 100~MIPs {\em and} the TCMT energy is less than 
50~MIPs.  These cuts vary linearly with energy, 
so that with a  30~GeV beam, for example, 
they are 329, 114 and 64~MIPs respectively. 
Pions which do not interact in the ECAL are retained in the sample, 
to permit comparison of the interaction cross-section between data and simulation.     
The fraction of events removed as muons varies with beam energy, but it is typically
$\sim 5\%$.

Double beam particles are very infrequent with the running conditions used in 2007, but 
in order to safely reduce any possible contribution from double beam particles, a cut is 
imposed, requiring that the total energy recorded in the ECAL+HCAL be less than 1.5 
times the beam energy.  Fewer than $0.5\%$ of events fail this cut.
In order to reduce the influence of interactions upstream
of the calorimeters, we remove events in which more than 50~MIPs are recorded in either of the 
first two layers of the ECAL.  This cut removes $\sim 2.5\%$ of data, 
with only a weak dependence on beam energy.  
A cut against beam halo is also applied, 
requiring the shower centroid to lie within $\pm50$~mm in both $x$ and $y$, 
corresponding approximately to the trigger acceptance; 
this cut also serves to remove a small number of 
showers close to the calorimeter edge, which are especially prone to lateral leakage.
This cut removes $\sim 5\%$ of data at the lowest energy and $\sim 1\%$ at the highest energy, 
where the beam is narrowest.   All of the cuts described so far are applied equally to data and 
Monte Carlo events.

\begin{figure}[tbp]
\centering \includegraphics[width=0.48\textwidth]{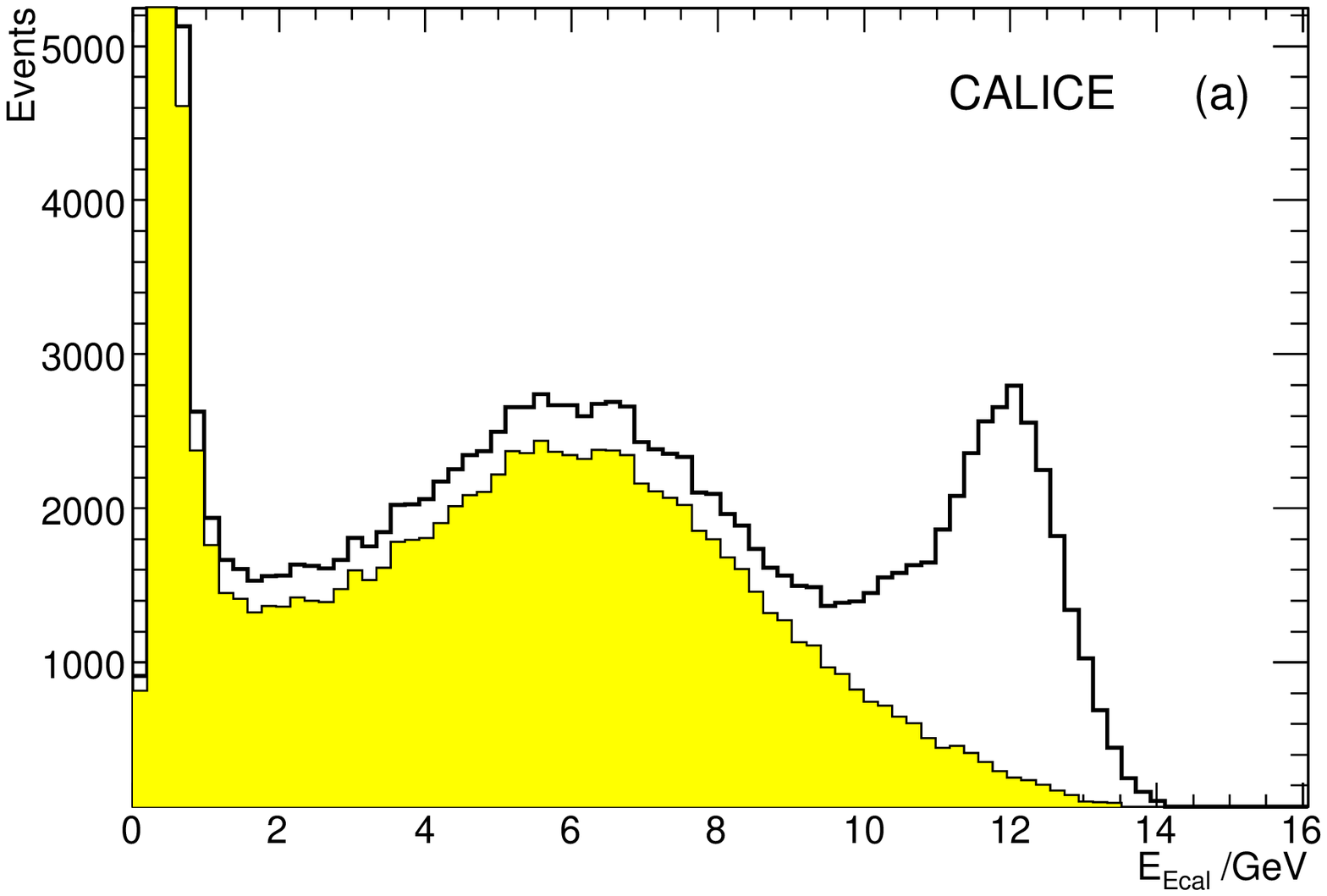}
 \includegraphics[width=0.48\textwidth]{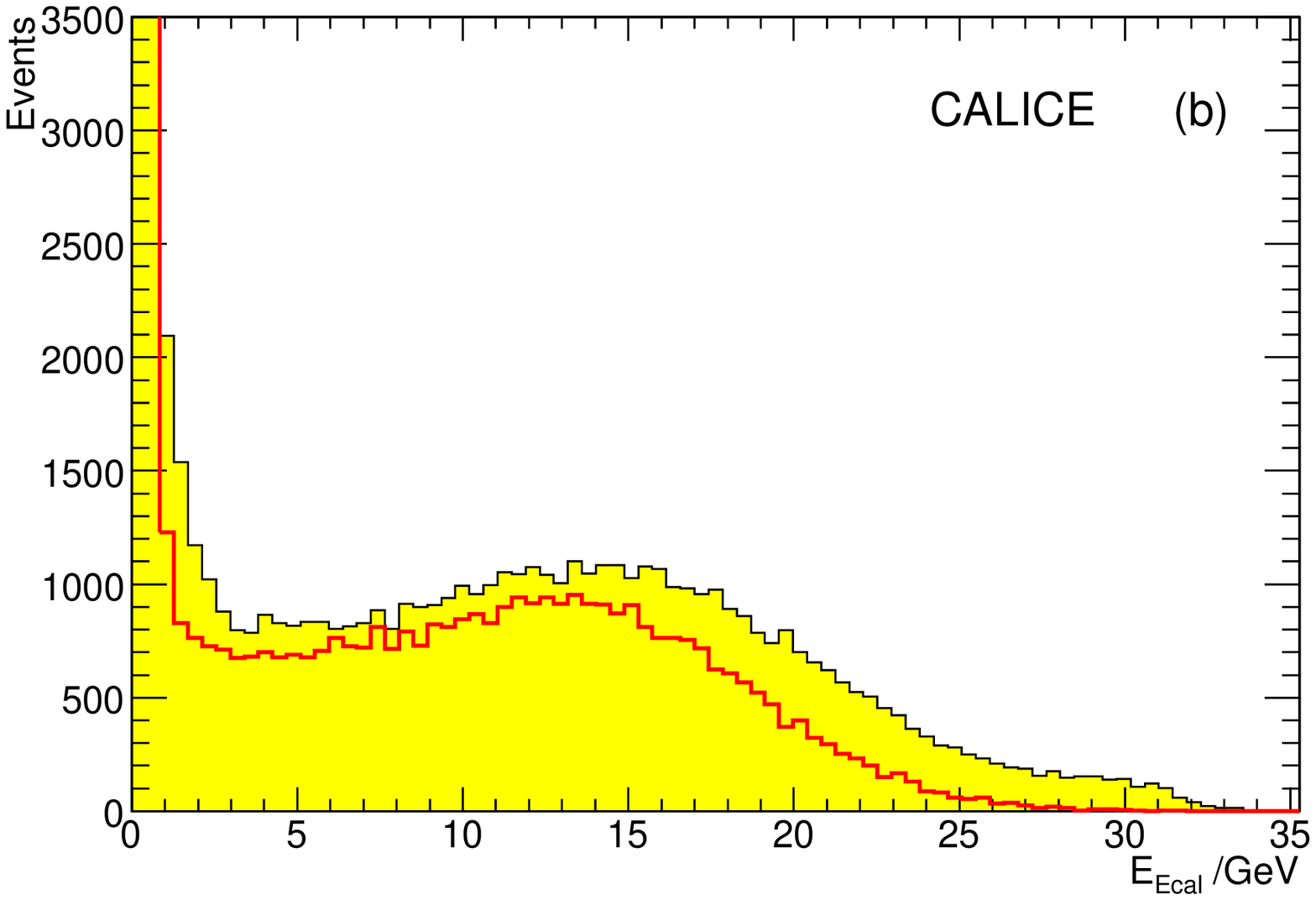}
\caption{\label{fig:Cerenkov}\em Energy observed in the ECAL for events with and without a 
signal from the beam \v{C}erenkov counter.  For this plot, the observed energy was 
converted from MIPS to GeV using a nominal conversion factor of 240 MIPs/GeV. (a) In the case of 
the 12~GeV negatively charged beam, the open region shows the total sample, and the
shaded region the contribution with no \v{C}erenkov signal. (b) For
the 30~GeV positively charged beam  the shaded histogram shows triggers with 
a \v{C}erenkov signal, and the open (red) histogram shows those with no \v{C}erenkov signal.}
\end{figure}

Electrons can provide a significant background to the pions in the negatively charged beams, 
and likewise protons are present in the higher energy positively charged beams.  
Their effect may be reduced by use of the \v{C}erenkov counter.  
In figure~\ref{fig:Cerenkov}(a) we compare, for typical negatively and positively charged beam runs, 
the distribution of the energy in the ECAL for triggers with and without a \v{C}erenkov signal.  
In the case of the negatively charged beam we see a clear peak at the full beam energy in 
the events with a \v{C}erenkov signal, which can be ascribed to electrons depositing their full 
energy in the ECAL.  Therefore, for the negatively charged beams we demand that the \v{C}erenkov have a signal.
Conversely, for the positively charged beams used here, the \v{C}erenkov pressure was set so that 
pions (and electrons) would yield a signal, while the heavier species would not. 
The ECAL energy distributions with and without a \v{C}erenkov signal are shown in figure~\ref{fig:Cerenkov}(b);
we see that for the events with no \v{C}erenkov signal, mainly protons,  
the distribution is shifted to lower values, 
and accordingly for these runs we demand that there be {\em no} \v{C}erenkov signal.
We also note that the contribution of positrons, which would show up as a peak
close to 30 GeV in the sample of events with a \v{C}erenkov signal, is small\footnote{ 
From fitting the region around the full beam energy, we 
estimate a positron contribution of $\sim$1\% at 30~GeV, decreasing to 0.4\% at 50~GeV and 0.1\% at 80~GeV.
This has a negligible impact on the results shown in the present paper.   
No perceptible electron contribution is seen in the negatively charged beam samples.}.  

In our previous study of electron showers in the ECAL~\cite{ref:ECALresp}, 
it was necessary to remove showers close to the edges of wafers, for two reasons.
Firstly, the guard rings lead to a significant dead region (2~mm wide), which can lead to 
a significant fraction of the electron energy not being recorded.  Secondly, we observed
an effect of 
correlated crosstalk between the charge deposited in the guard rings and the peripheral 
cells of a wafer, whereby a square pattern of hits was sometimes 
seen in the core of a high energy shower~\cite{ref:ECALcomm}.  
The first of these problems is less important 
for hadronic showers because of their greater transverse width, and in any case the effect 
is sufficiently well simulated.  The second problem is not seen in hadronic showers, because 
the energy densities in the core of the showers are much smaller.     
Therefore no further cuts on shower position are imposed.

\section{Comparison between data and simulation}
\label{sect:Comp} 
\subsection{Total ECAL energy}
\label{sect:CompEecal}
\begin{figure}[tbp]
\centering \includegraphics[width=0.95\textwidth]{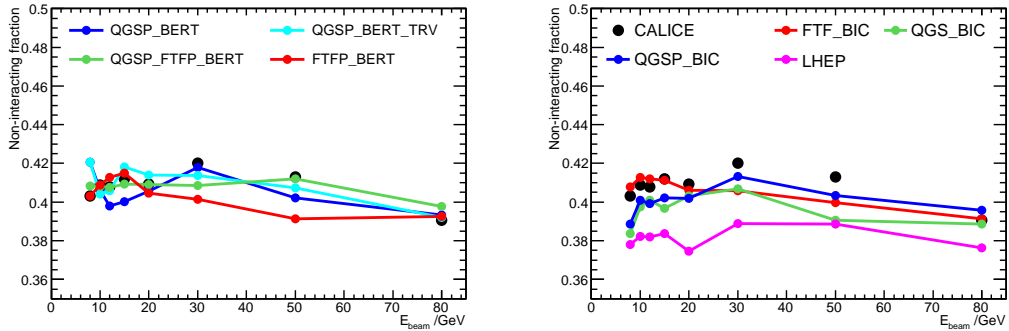}
\caption{\label{fig:fNonInt}\em Mean fraction of non-interacting pions (energy $<$100 MIPs) in the ECAL, 
plotted  as a function of beam energy.  For this purpose, the three ECAL stacks were combined 
with weights 1:1:1.  The data are compared with the predictions of simulations using different 
{\tt GEANT4} physics lists.  The models are separated into two plots in the interests of clarity, 
with the physics lists which incorporate the Bertini model on the left, and the others on the right. }
\end{figure}

The total energy recorded in the ECAL is a useful starting point for comparison between data 
and simulation.  The event selection cuts were designed to retain pion events wherever 
they interacted in the calorimeter system.  A significant fraction of pions should not start to
shower in the ECAL.  These events are characterised by a MIP-like energy in all layers 
(apart from occasional $\delta$-ray emission) and accordingly the ECAL energy shows 
a large peak at $\sim$50~MIPs.  The fraction of such events can be used to test the interaction 
cross-sections in {\tt GEANT4}.  In figure~\ref{fig:fNonInt} we show the fraction of selected pions 
depositing fewer than 100~MIPs in the ECAL, as a function of energy, compared with the simulations. 
Most of the models give a good description of the fraction of non-interacting pions at all energies, 
agreeing with data within 0.01-0.02.  The {\tt LHEP} physics list is the most discrepant.  
This gives confidence in the cross-sections simulated in {\tt GEANT4}.  It could also be interpreted as 
an indication that any residual beam contamination by kaons or (anti-)protons is  
small, since these species would be expected to have different interaction cross-sections.   
 
\begin{figure}[tbp]
\centering \includegraphics[width=0.45\textwidth]{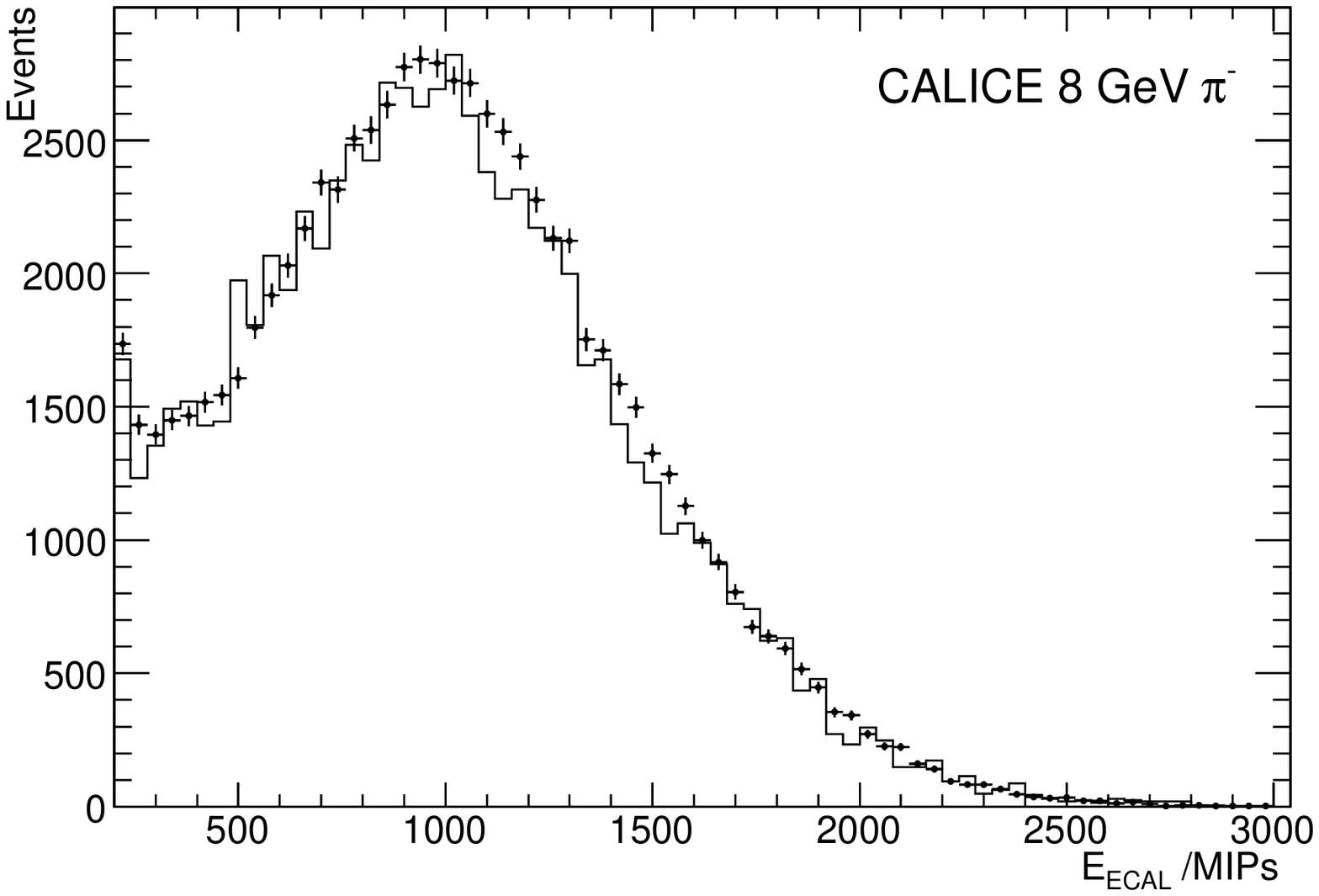}
\centering \includegraphics[width=0.45\textwidth]{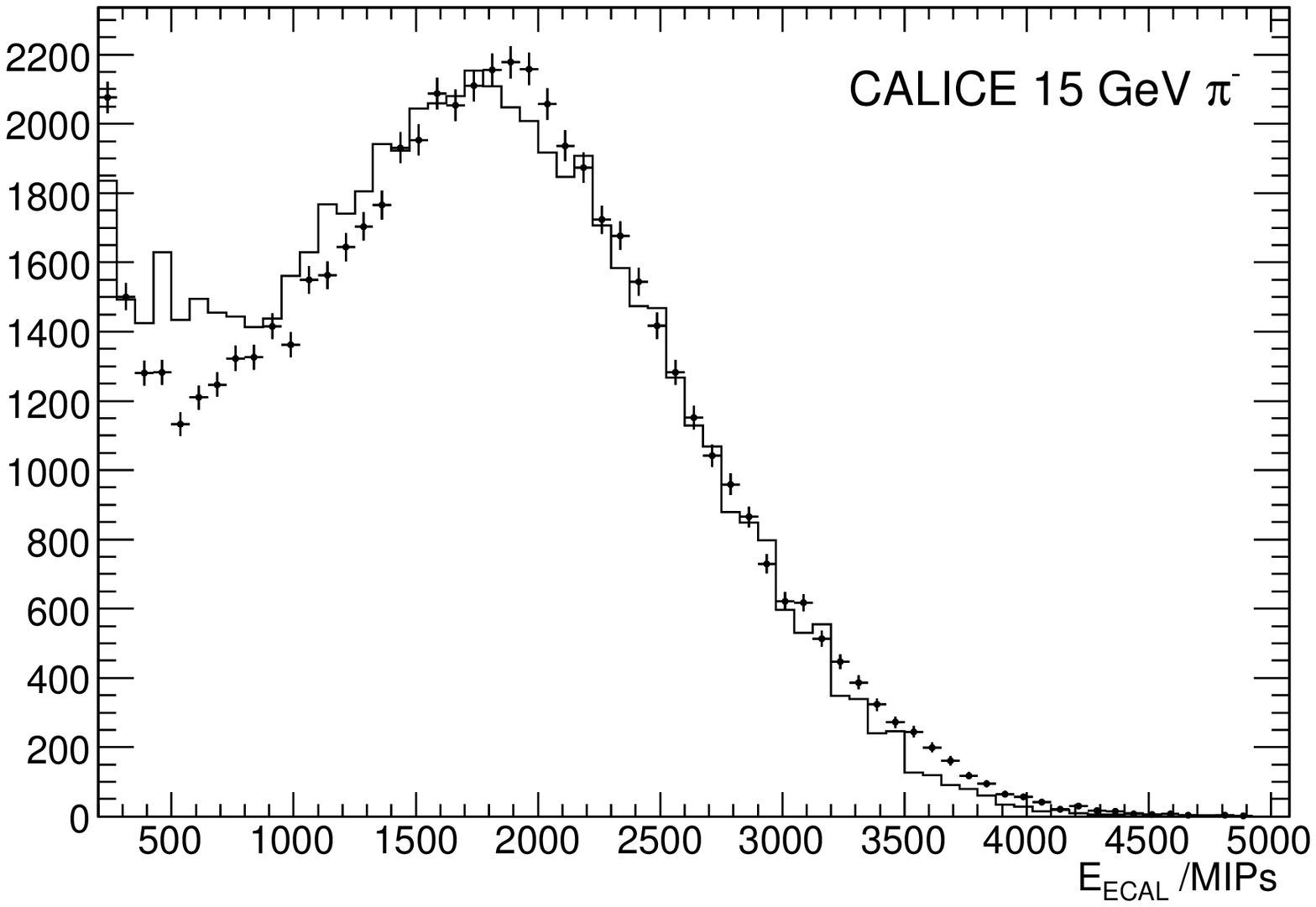}
\newline
\centering \includegraphics[width=0.45\textwidth]{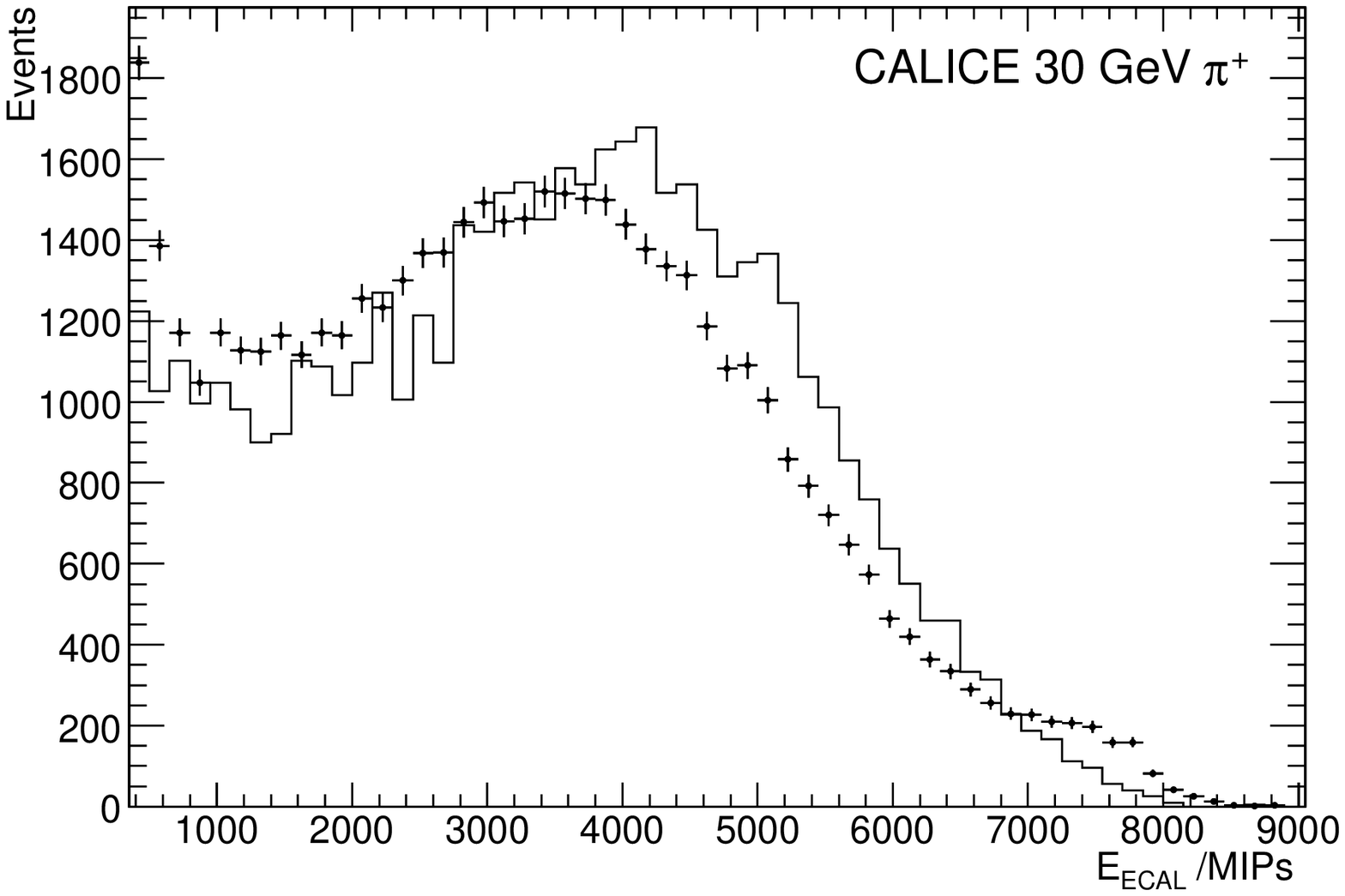}
\centering \includegraphics[width=0.45\textwidth]{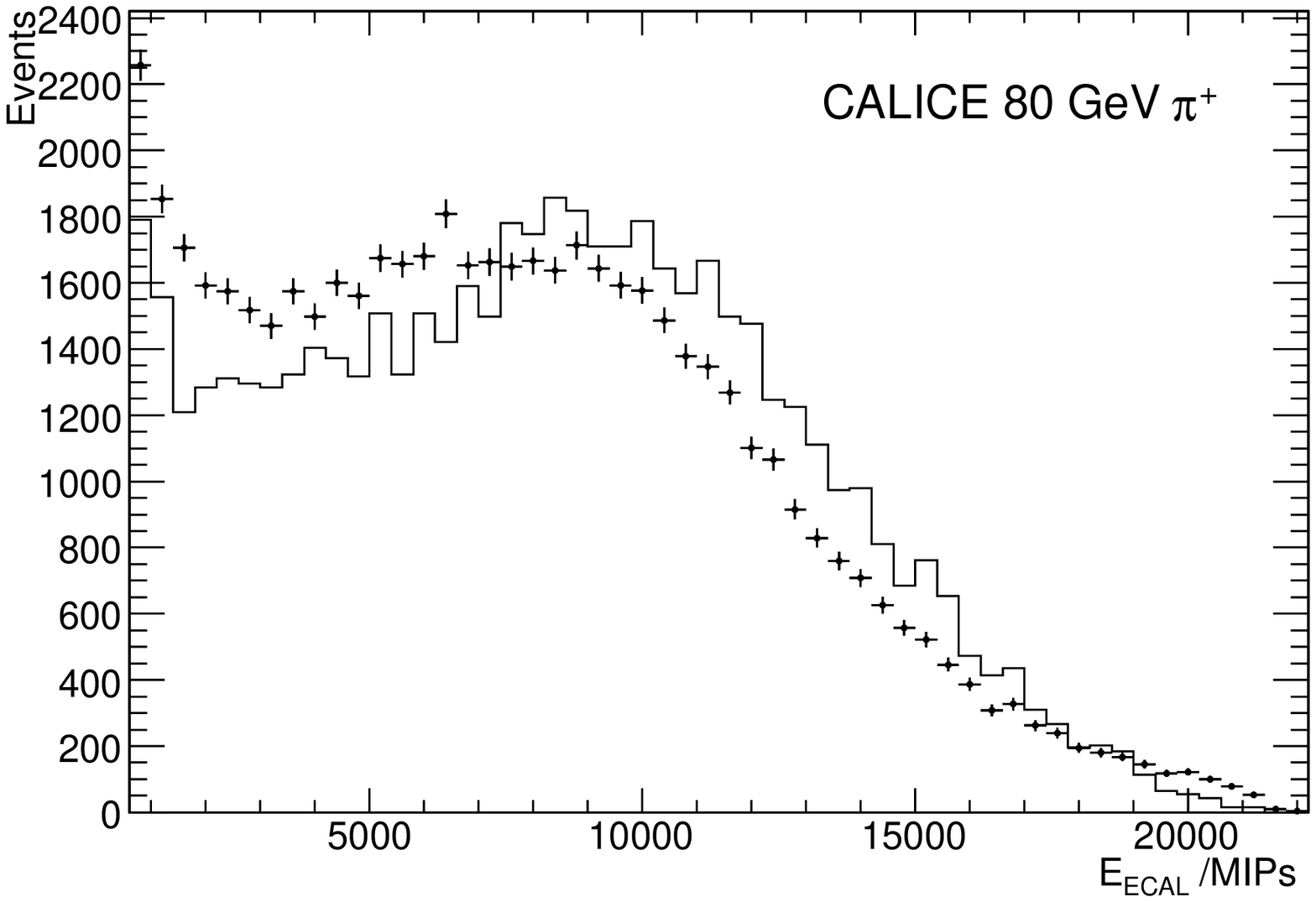}
\caption{\label{fig:Etot}\em  Distributions of total energy recorded in the ECAL at 8, 15, 30 and 80~GeV
(points with error bars), 
compared with Monte Carlo predictions using the QGSP\_BERT physics list (solid histograms). 
The distributions are normalised to the same numbers of selected events
(including the non-interacting peak). }
\end{figure}

We now consider the energy deposited in the ECAL by those pions which have their 
first interaction in the ECAL.  
In the study of electromagnetic showers~\cite{ref:ECALresp}, it was appropriate 
to combine the energies in the three stacks of the ECAL taking account of their
relative sampling fractions.  Since the thicknesses of tungsten in the three stacks 
lie in the ratio 1:2:3, the weighting of the energy should be in roughly these 
ratios.  It is less obvious that this is the correct procedure in the case of hadronic 
showers, but for the purposes of the present comparison, we choose to combine energies of
each stack using the na\"{\i}ve weighting factors of 1, 2 and 3.  
The level of agreement between data and simulation 
is not found to be sensitive to this choice.  
In figure~\ref{fig:Etot} we compare the distribution of  recorded energy in data 
with simulation using the {\tt QGSP\_BERT} physics list, at four typical energies.
The non-interacting peak at low energies has been suppressed.  
The broad peaks in the distributions in figure~\ref{fig:Etot} represent pions which 
started to shower in the ECAL, extending roughly to a point corresponding to all the 
energy being deposited in the ECAL (approximately 250~MIPs per GeV).  The main peak is quite well
modelled at 8 and 15~GeV, while at 30 and especially 80~GeV the Monte Carlo predicts significantly 
more energy than observed in the data.\footnote{
In ref.~\cite{ref:ECALresp} the energy response of the ECAL for electrons
was seen to be linear up to  45~GeV.  Since the energy densities in electromagnetic 
showers are significantly greater than those encountered in hadronic showers, we can be confident that 
saturation effects will not play any significant part in the hadron beam data.}

\begin{figure}[tbp]
\centering \includegraphics[width=0.95\textwidth]{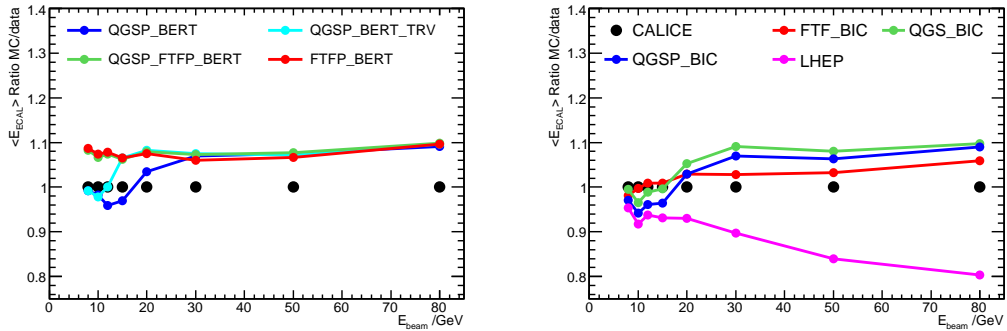}
\caption{\label{fig:EtotAll}\em Ratio of simulation to data for the 
mean energy recorded in the ECAL, plotted  as a function of beam energy.  
The data are compared with the predictions of simulations using different {\tt GEANT4} physics lists. }
\end{figure}

In order to incorporate all models and energies into the analysis, 
a useful global measure is the average energy deposited in the ECAL, i.e. the mean values of the
histograms such as those shown in figure~\ref{fig:Etot}, excluding the non-interacting pions.  
This can be computed for data and for all the {\tt GEANT4} physics lists under consideration, 
and is plotted in 
figure~\ref{fig:EtotAll} in the form of ratios of simulation to data.  At 8~GeV, all of the 
models lie within 10\% of the data, and most within 5\%.   
As energy increases, the LHEP model falls steadily further below the data.  
The other models either remain $\sim5-10$\% above the data or increase so that 
at high energies, all of the models lie $\sim5-10$\% above the data. 
Overall, {\tt FTF\_BIC} is the most consistent with this aspect of the experimental data.

\subsection{Transverse energy profile}
\label{sect:CompTrans}

Good modelling of the transverse shower width is of importance for 
the development of particle flow algorithms, since it affects the degree of overlap between showers, 
and therefore the efficiency for separating them.  For each hit in the ECAL, we determine 
the transverse distance between the centre of the pad and the shower barycentre in $x$ and $y$.
By histogramming this radial distance, we form the transverse shower profile.   We weight the hits by
their energy\footnote{In this case 
we do not apply different weighting factors to each stack.}, to emphasise the flow of 
energy in the shower.  This also has the benefit of minimising 
possible residual effects of noisy cells.   

\begin{figure}[tbp]
\centering \includegraphics[width=0.45\textwidth]{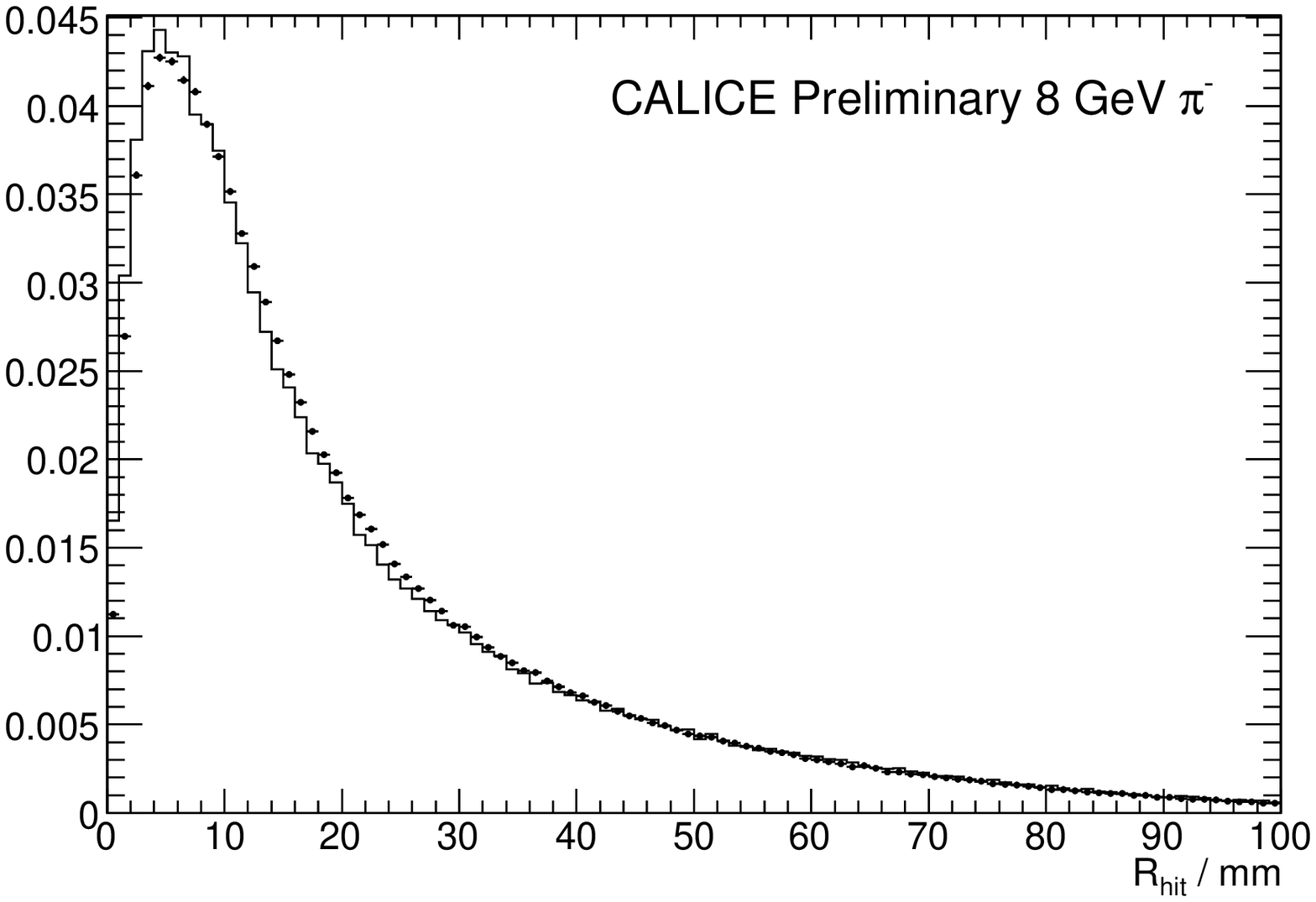}
           \includegraphics[width=0.45\textwidth]{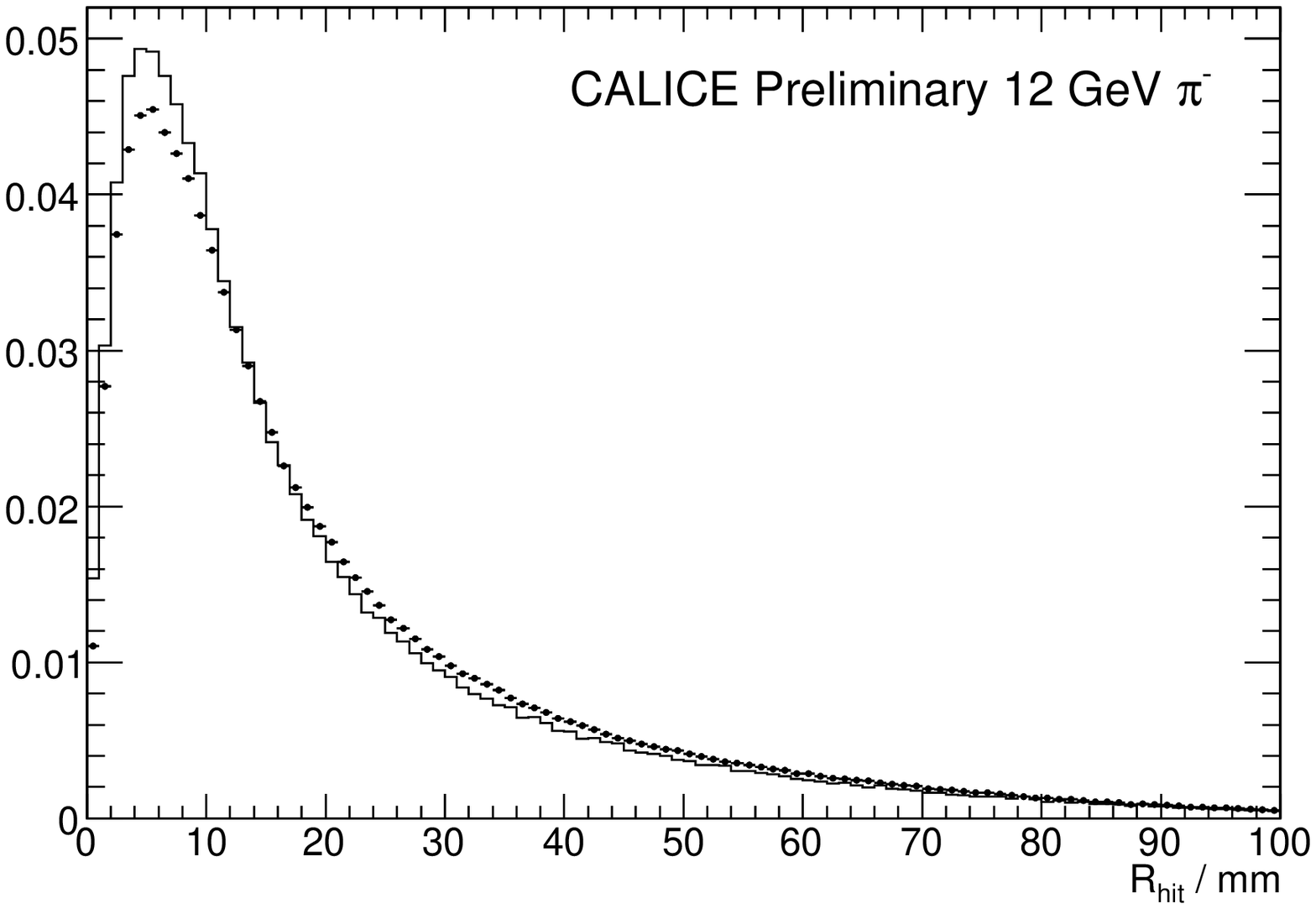} \\
           \includegraphics[width=0.45\textwidth]{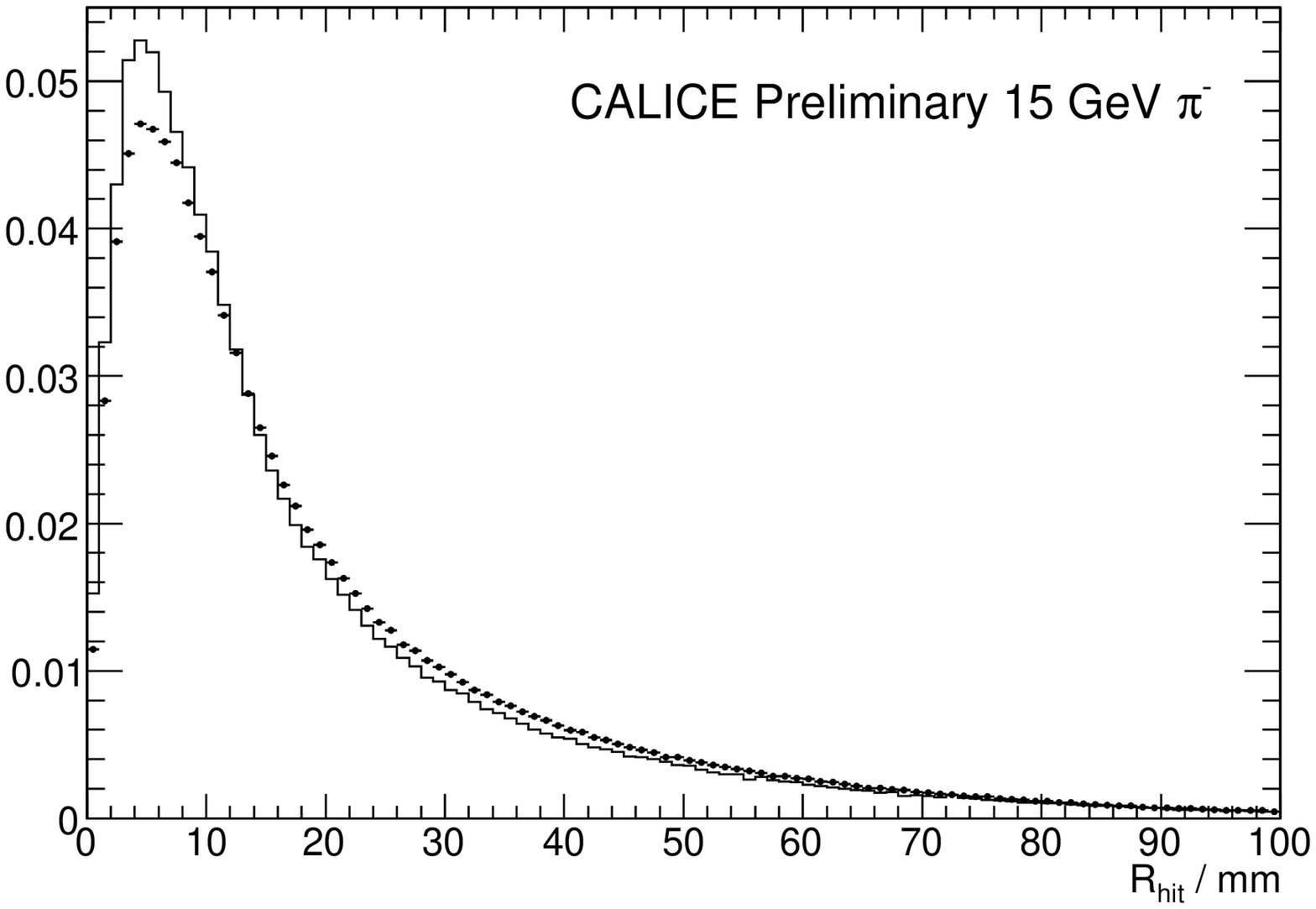} 
           \includegraphics[width=0.45\textwidth]{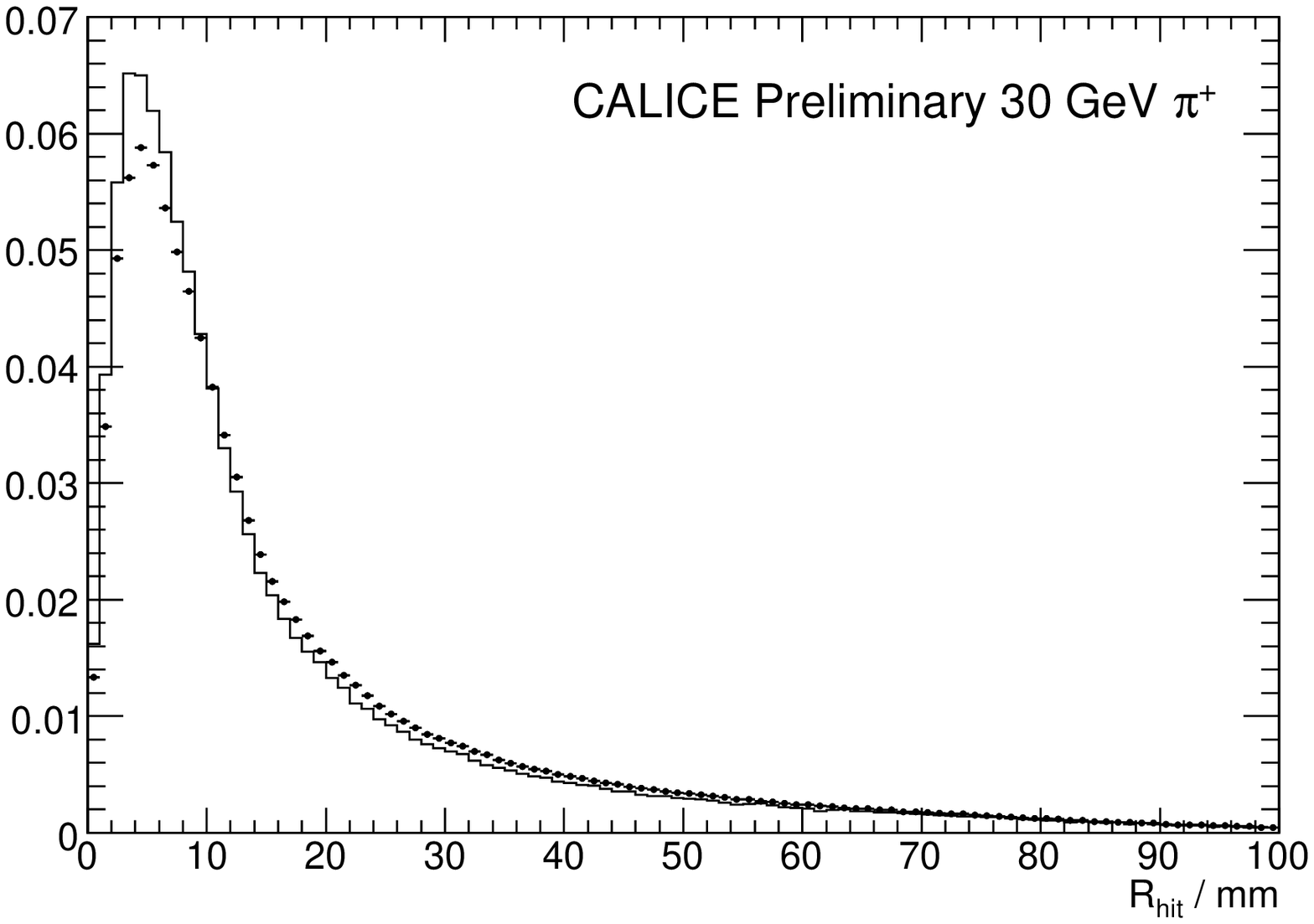}
\caption{\label{fig:Trans_QGSP_BERT}\em Radial distribution of hits (energy weighted) for data at 
four typical energies (points with errors) compared with Monte Carlo (solid histograms)
 using the {\tt QGSP\_BERT} physics list. The distributions are normalised to unity. }
\end{figure}

In figure~\ref{fig:Trans_QGSP_BERT} we show these transverse energy
profiles for four typical energies, compared with the simulation based on the QGSP\_BERT physics list.
The entries in these plots are weighted by energy, 
and we want to focus here on the shape rather than the
normalisation, which is essentially governed by the differences 
in energy response which were already considered in figure~\ref{fig:EtotAll}.
Therefore in figure~\ref{fig:Trans_QGSP_BERT} 
the distributions are normalised to unity, in order to facilitate the comparison of their shapes.     
We see that the data at 8~GeV are
quite well modelled, while at higher energies the data tend to lie below the simulation at 
small radii, and above at higher radii; in other words the simulation underestimates the width 
of that part of the shower which is contained in the ECAL. 

\begin{figure}[tbp]
\centering \includegraphics[width=0.95\textwidth]{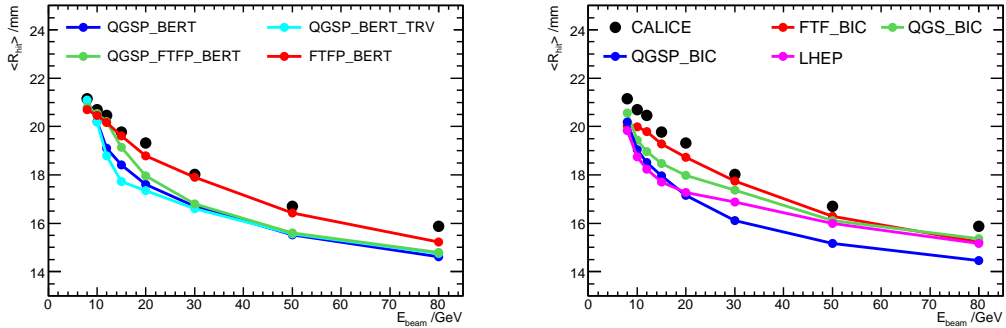}
\caption{\label{fig:TransMean}\em  Mean energy-weighted shower radius in the ECAL as a function of beam energy.  
The data are compared with the predictions of simulations using different {\tt GEANT4} physics lists. }
\end{figure}

In order conveniently to compare all models and energies, in figure~\ref{fig:TransMean} we show the 
mean energy-weighted shower radius (i.e. the mean values 
of distributions such as  figure~\ref{fig:Trans_QGSP_BERT})
as a function of energy.  The observed shower becomes narrower with increasing energy, 
both in data and in all of the models.  All of the physics lists underestimate the shower width 
at almost all energies, typically by around 10\%. 
The models {\tt FTF\_BIC} and especially {\tt FTFP\_BERT} tend to lie closest to the data, 
especially at intermediate energies.\footnote{For comparison, the mean radius of 
electron-induced showers, selected using the methods of ref.~\cite{ref:ECALresp}, 
shows agreement between data and simulation at the $\sim1-2\%$ or 0.1--0.2~mm level.}
     
\begin{figure}[tbp]
\centering \includegraphics[width=0.95\textwidth]{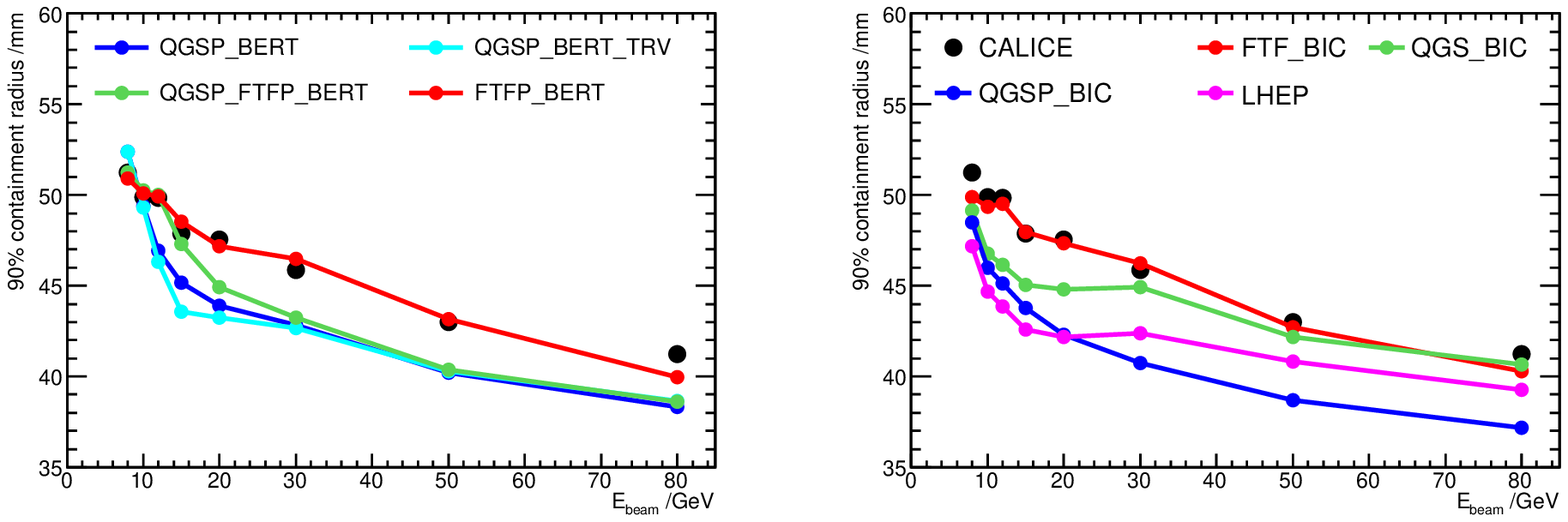} 
\\
\includegraphics[width=0.95\textwidth]{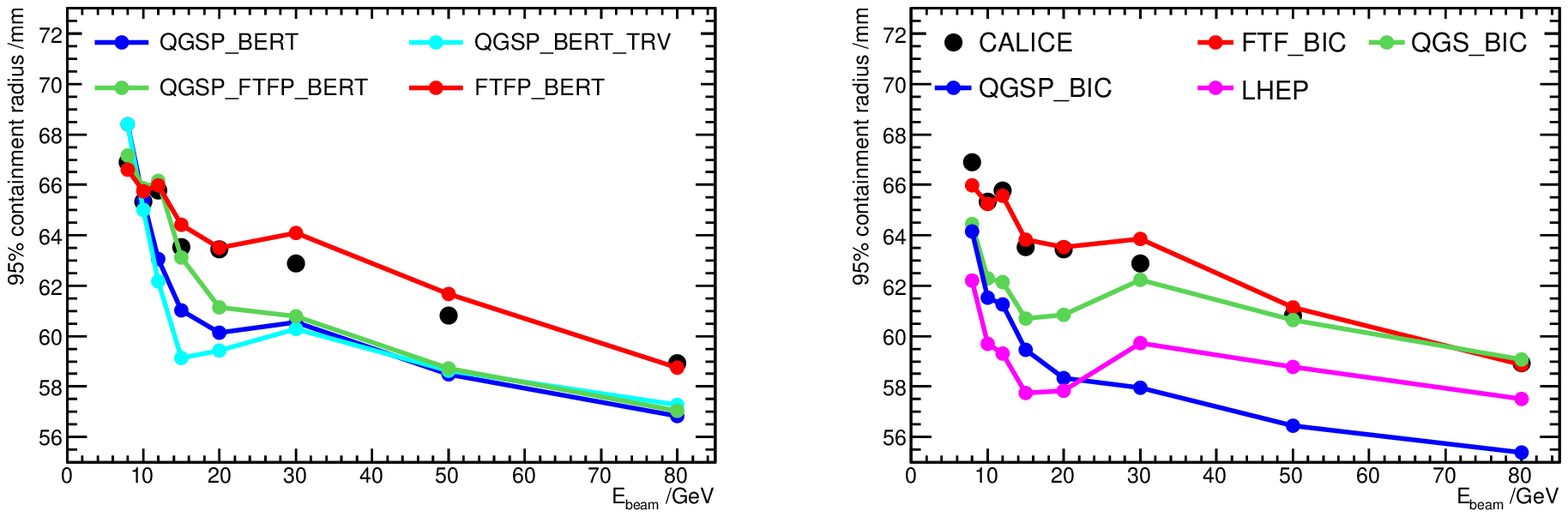}
\caption{\label{fig:R90}\em  Radii required to contain 90\% (upper two plots) or 95\% (lower two plots) of the energy seen in the ECAL, as a function of beam energy.  
Data are compared with simulation for various physics lists.}
\end{figure}

Of course, the mean shower radius is only one measure of the transverse shower profile.
In figure~\ref{fig:R90} we focus on the tails of the showers by plotting the radii needed to 
contain  90\% or 95\% of the observed ECAL shower energy.   As before, we find that most of the 
models tend to underestimate the data, and none of them gives a really satisfactory description.  
Again, the most successful physics lists are clearly {\tt FTFP\_BERT} and {\tt FTF\_BIC}. 
It should be stressed that these observations refer only to that part of the hadronic shower which is 
detected in the ECAL, i.e.\ roughly the first interaction length --- the transverse tails are 
not fully sampled, and nor is most of the longitudinal tail of the shower.   

\subsection{Longitudinal distribution of interaction point}
\label{sect:CompIntLayer}

\begin{figure}[tbp]
\centering \includegraphics[width=0.48\textwidth]{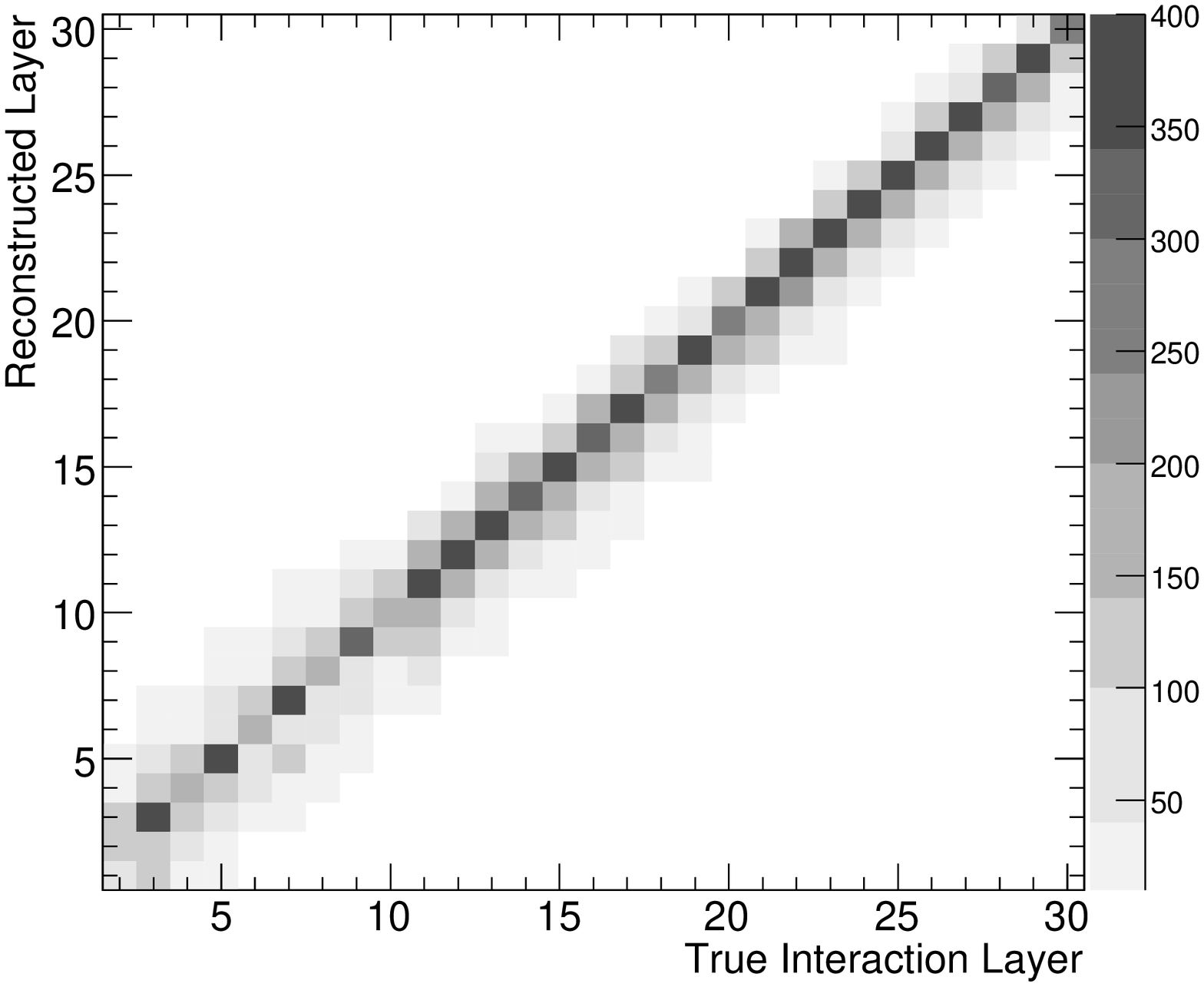}
\includegraphics[width=0.48\textwidth]{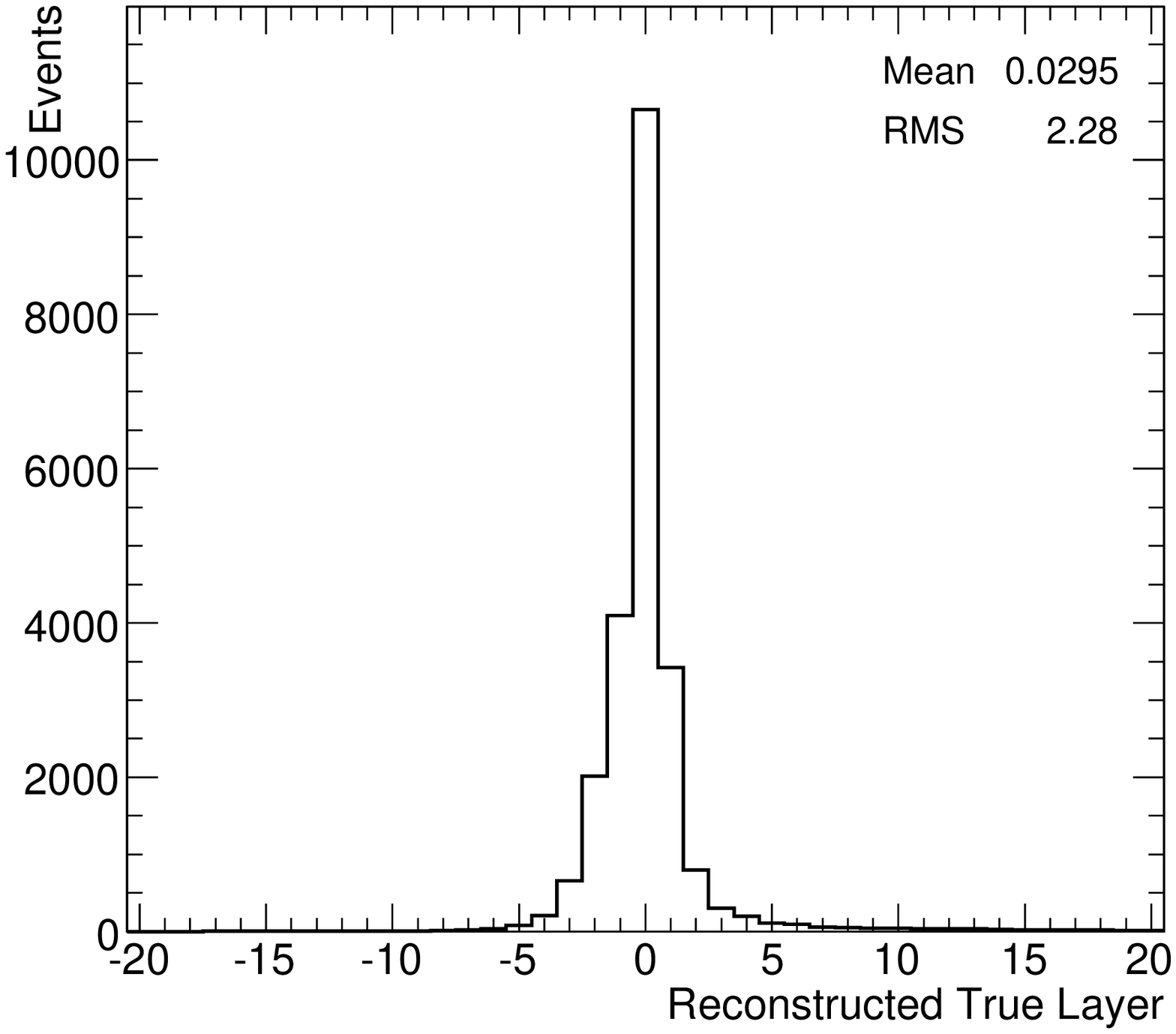}
\caption{\label{fig:InteractionLayer}\em Left: comparison between reconstruction and truth for the layer
identified as the interaction layer by the algorithm described in the text.  
This example corresponds to 
a 20~GeV $\pi^-$ beam simulation, using the QGSP\_BERT physics list.  
The grey scale indicates the number of events in each bin, 
where bins containing fewer than 10 events (amongst a total of more than 23000) have been suppressed 
for the sake of clarity. Right: distribution of (reconstructed $-$ true) layer for this sample.}
\end{figure}

The fine granularity of the CALICE ECAL makes it possible to identify the point at which the incident 
pion makes its primary interaction, and this will be used in some of the results shown below.  
In general one expects to see a track of MIP-like hits 
until the interaction point, followed by some multiparticle shower structure thereafter. 
The algorithm used for the present analysis has the merit of simplicity.  
Firstly, the energy in each layer 
is computed, after excluding isolated hits
(those with no neighbours in adjacent cells in the same layer).  
The interaction layer is defined as the first layer containing at least 10 MIPs of energy, 
provided that at least two of the following three layers also show energy greater than 10 MIPs.    
In simulated events, the true interaction point is stored using information about the true activity
in the Monte Carlo.  In figure~\ref{fig:InteractionLayer} we show the correlation between 
the true interaction layer and that found by our algorithm, for a typical energy 
and Monte Carlo physics list.  The correct layer is identified
within $\pm 1$ in $\sim$70\% and within $\pm2$ in almost 90\% of events.  
The mean of the difference between 
the true and reconstructed layer lies within $\pm1$ layer for 
all physics lists and energies studied here.  

\begin{figure}[tbp]
\centering \includegraphics[width=0.48\textwidth]{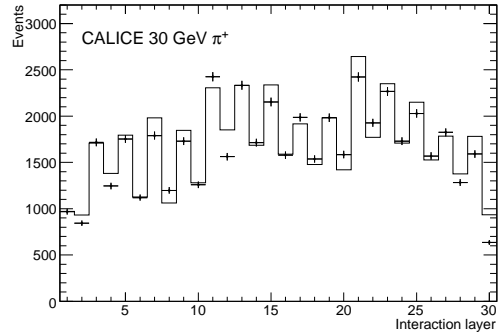}
\caption{\label{fig:InteractionLayer2}\em  Distribution of the reconstructed 
interaction layer in the  ECAL for 30~GeV data (points), 
compared with Monte Carlo predictions using the QGSP\_BERT physics 
list (solid histogram). }
\end{figure}

In figure~\ref{fig:InteractionLayer2} we show the distribution of the
reconstructed interaction layer for 30~GeV data compared with a typical Monte Carlo
run. The alternation between odd and even layers reflects the additional material 
associated with PCBs and the carbon-epoxy support structure; passing from an odd-numbered 
layer to an even layer, a particle will encounter just a tungsten plate, while going from an 
even to an odd layer it will also pass through two PCBs and part of the 
mechanical support structure, adding an additional $\sim0.008\lambda_{\mathrm{int.}}$.
The attenuation of the beam through each stack is clearly seen, becoming stronger in 
later stacks as the tungsten thickness increases.  The upward steps 
at layers 10 and 20 result from the increase of thickness of the tungsten plates at these points. 
The lower values in layers 1 and 2 are a by-product of the cut designed 
to remove upstream interactions.  These features are all seen in the simulation, and are generally 
well modelled.  The following analysis concentrates on interactions in the first 
stack, for which the agreement is particularly good.
This distribution is similarly well described at all energies
by all physics lists, which suggests that there is no significant problem 
with the high energy cross-sections on tungsten in this version of {\tt GEANT4}.  
This is therefore not a directly useful distribution for 
discriminating between models.

\subsection{Longitudinal energy profile}
\label{sect:CompLong}
\begin{figure}[tbp]
\centering \includegraphics[width=0.48\textwidth]{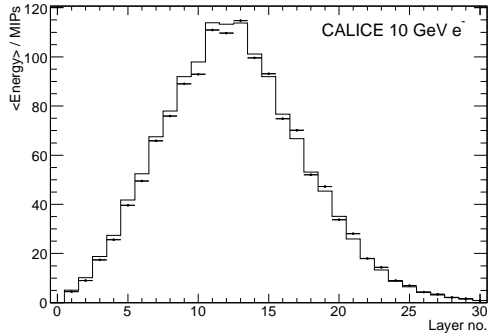}
\caption{\label{fig:Long_e}\em  Energy per layer in the ECAL for 10~GeV electron data (points), 
compared with Monte Carlo predictions (solid histogram). }
\end{figure}

It is of interest to study the longitudinal development of the showers.  
In the case of electrons,  we find that the mean depth of 
electron-induced showers, selected using the methods of ref.~\cite{ref:ECALresp}, 
shows good agreement between data and simulation, to better than $0.1X_0$ 
(corresponding to less than 25\% of a 1.4~mm tungsten layer), and the standard 
deviation is modelled to better than $0.03X_0$. 
A typical comparison is shown in figure~\ref{fig:Long_e}.  We therefore believe that the 
material content of the ECAL is well modelled.  
The case of hadronic showers is, however, complicated by the different points in the ECAL 
at which the shower is initiated.  The observed longitudinal distribution 
of energy in the ECAL is the convolution of the intrinsic shape of showers with
the distribution of shower starting points.  In order to have some sensitivity
to the composition of the shower in terms of different particle species, we would
like to measure the shower profile with respect to the primary interaction point.

In Sect.~\ref{sect:CompIntLayer} we explained our algorithm for identifying
the layer closest to the interaction point, and showed that it was reasonably
reliable and well modelled.  We now proceed to compute the shower profile in terms of layers
after the interaction layer.  This is made more complicated by the different 
sampling fractions in the three ECAL stacks. We circumvent this by the following 
expedient.  For hits in the first stack (tungsten thickness 1.4~mm), 
we simply use the measured energy in each layer.  In the second stack 
(tungsten thickness 2.8~mm), in addition to the energy measured in each layer, we introduce
an additional fictitious {\em pseudolayer} midway between each physical layer, whose energy is 
estimated by linear interpolation between the layers on either side.  Similarly, in the
third stack (tungsten thickness 4.2~mm) we introduce two pseudolayers between each physical
layer.  In this way, the energy deposition is estimated in 60 layers each separated by 
an effective separation of 1.4~mm of tungsten.  The longitudinal 
shower profile (i.e. the mean energy, in MIPs, per layer) is then trivially 
computed starting at the interaction layer measuring the depth in units of these 
effective 1.4~mm layers.  This calculation will lead to some non-trivial correlations 
between neighbouring bins in the longitudinal profile.  It also neglects differences
in the material between successive samplings caused by other passive materials such as PCBs,
though this effect will be averaged out by combining different interaction layers. 
Furthermore, both these features will also be present in the simulation.
For this study, we restrict the interaction layer to lie in the first stack, so that 
at least 50 effective layers are available for observation the shower profile, corresponding to
$\sim20\;X_0$ or $0.8\;\lambda_{\mathrm{int.}}$.       
   
In the case of the {\tt GEANT4}/Mokka simulation, it is possible to store a breakdown of the 
fractions of the energy of each hit contributed by various particle species.
In this way, one can break down the longitudinal shower profile into its constituent parts.
We choose to separate the energies deposited by 
electrons, positrons, ``mesons'' (charged pions, kaons and muons),
protons (including antiprotons) and ``others'' (e.g.\ deuterons, $\alpha$-particles,
hyperons).  

\begin{figure}[tbp]
\centering \includegraphics[width=0.98\textwidth]{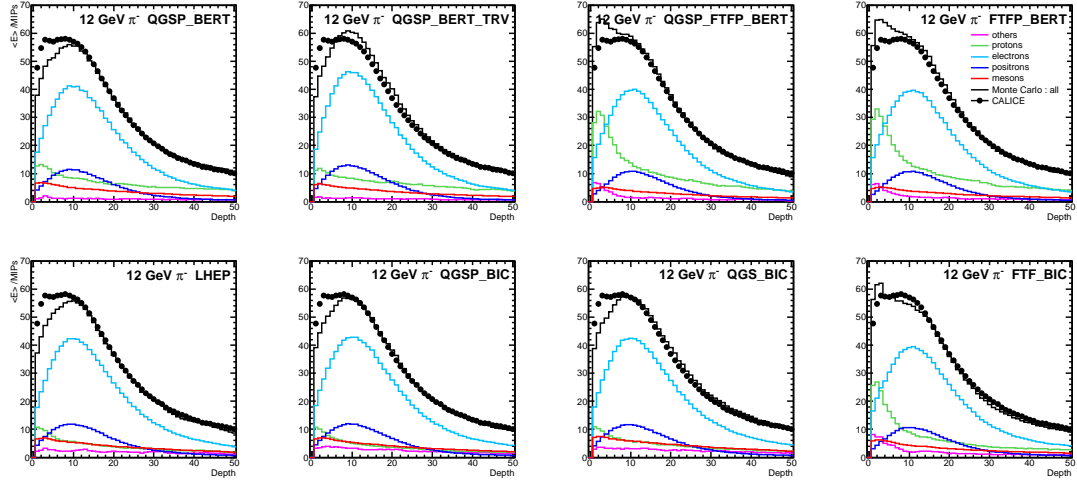}
\caption{\label{fig:Long12}\em Longitudinal energy profiles for 12~GeV $\pi^-$ 
data (shown as points), compared with simulations using different physics lists. 
The mean energy in MIPs is plotted against
the depth after the initial interaction, in units of effective 1.4~mm 
tungsten layers. The total depth shown corresponds to $\sim20\;X_0$ or $0.8\;\lambda_{\mathrm{int.}}$.
The breakdown of the Monte Carlo into the energy deposited by different particle categories
is also indicated.}
\end{figure}

In figure~\ref{fig:Long12} we show, for  12~GeV
$\pi^-$, the longitudinal energy profiles data compared with simulations using 
different physics lists.  In each case, we also show the simulation broken down into each of the above 
categories.  The ``mesons'' show a smooth slowly falling shape, while the electrons and positrons 
show the characteristic rise and fall of an electromagnetic shower, peaking after
$\sim$10-15 layers ($\sim4-6 X_0$). The electrons, however, also exhibit a long tail; this is because 
part of the ionisation energy loss of other particles is simulated in the form of 
discrete knock-on 
electrons, and therefore is included in the electron contribution.  The protons, and the
small contribution from ``others'', seem to exhibit two components -- 
a sharp peak in the first few layers which, because of its short range,  
can be attributed to low energy 
nuclear break-up fragments, and a long range tail similar to the mesons.
We note that there are significant differences between the models, most conspicuously in the  
short range proton component.  The comparison with data
demonstrates that our calorimeter has sufficient longitudinal 
granularity to be able to offer some discrimination between these components.  

\begin{figure}[tbp]
\centering \includegraphics[width=0.98\textwidth]{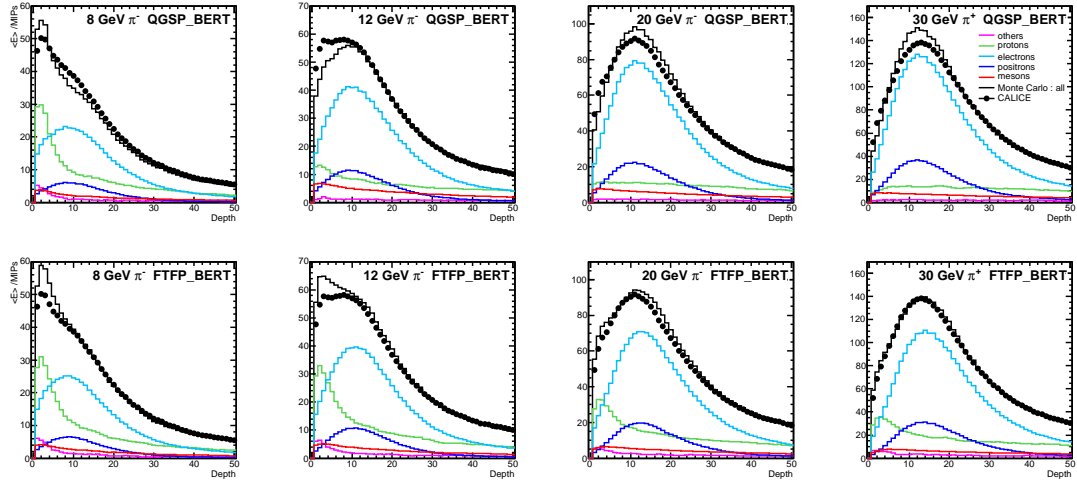}
\caption{\label{fig:LongModels}\em Longitudinal energy profiles for data  (shown as points)
compared with 
simulations using two physics lists,
{\tt QGSP\_BERT} and {\tt FTFP\_BERT}, at four typical energies. 
The breakdown of the Monte Carlo into the energy deposited by different particle categories
is also indicated.}
\end{figure}

The main features of the data distribution in figure~\ref{fig:Long12} may be summarised as follows:
\begin{itemize}
\item A small peak is seen in the first few layers, 
which we attribute to nuclear fragments, mainly protons.  
As noted above, the models exhibit considerable variation in 
their predictions in this region, associated with differences in their proton yields.
None agrees well with data, with the {\tt FTF}-based physics lists
 overestimating the data, and the others undershooting.     
\item A broad peak occurs around layer 10, which we ascribe to the electromagnetic component.  
This is tolerably well modelled by all physics lists, with some variations in normalisation.
\item A long tail follows, which appears to be generally well modelled. 
\end{itemize} 
In figure~\ref{fig:LongModels} we show similar distributions for two physics lists, 
{\tt FTFP\_BERT} and {\tt QGSP\_BERT}, at four typical energies.  Clearly both models have 
imperfections, but on balance {\tt FTFP\_BERT} is probably the more successful.  

\begin{figure}[tbp]
\centering \includegraphics[width=0.95\textwidth]{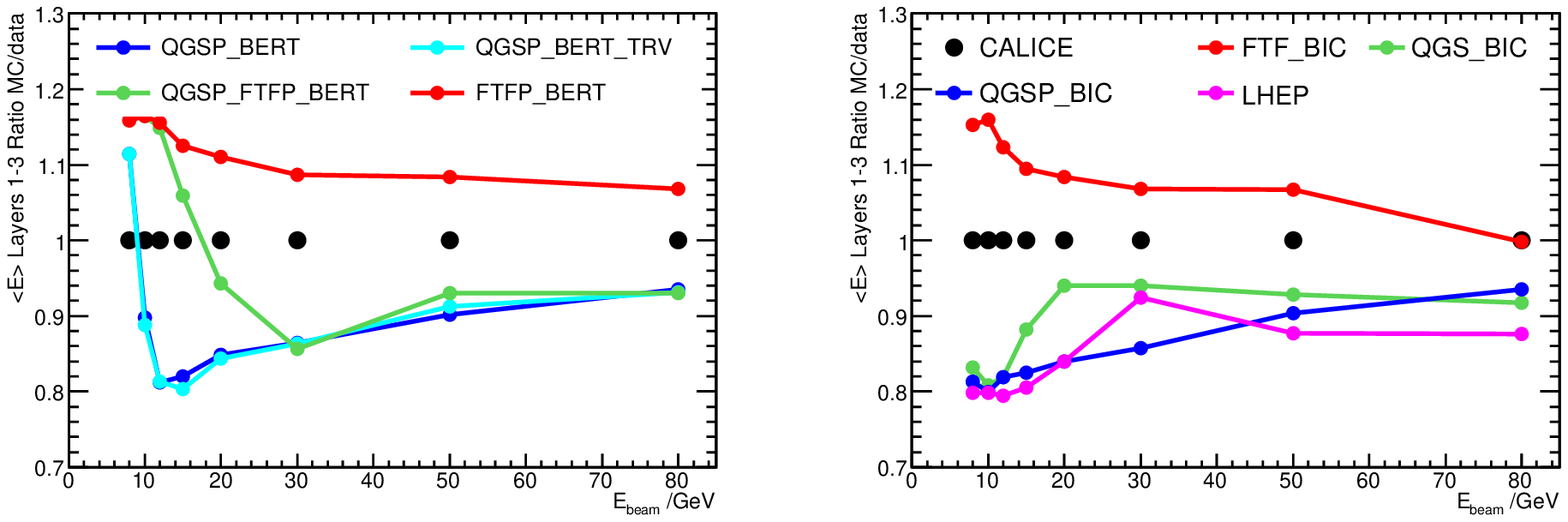}
\includegraphics[width=0.95\textwidth]{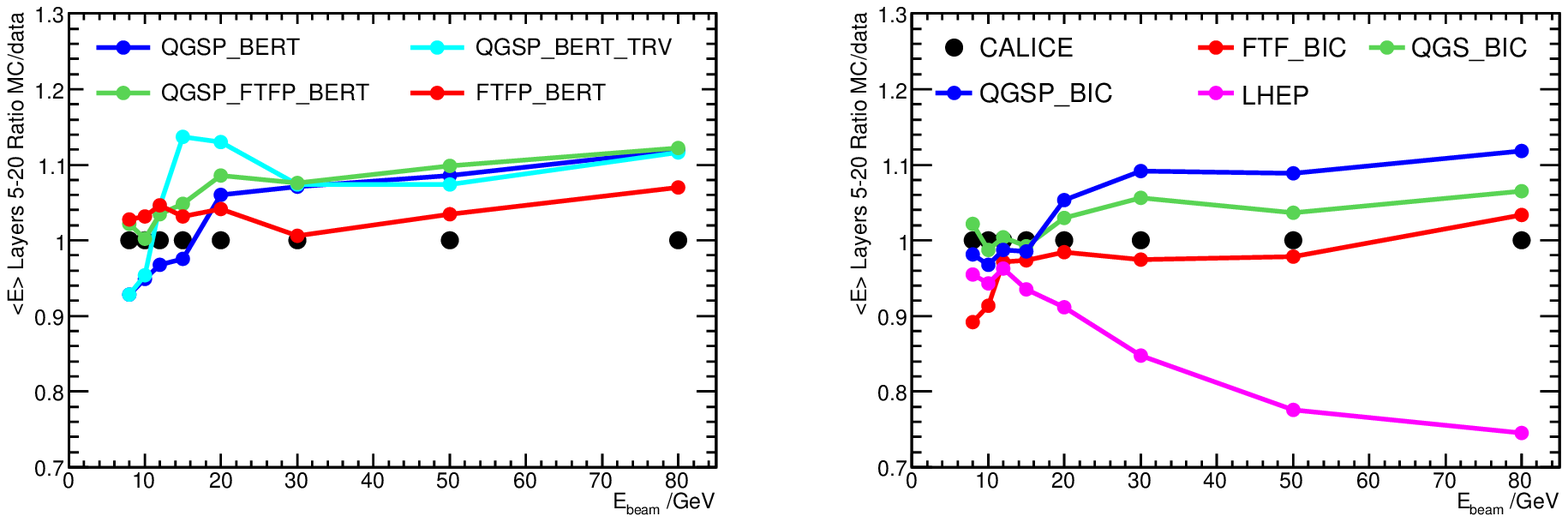}
\includegraphics[width=0.95\textwidth]{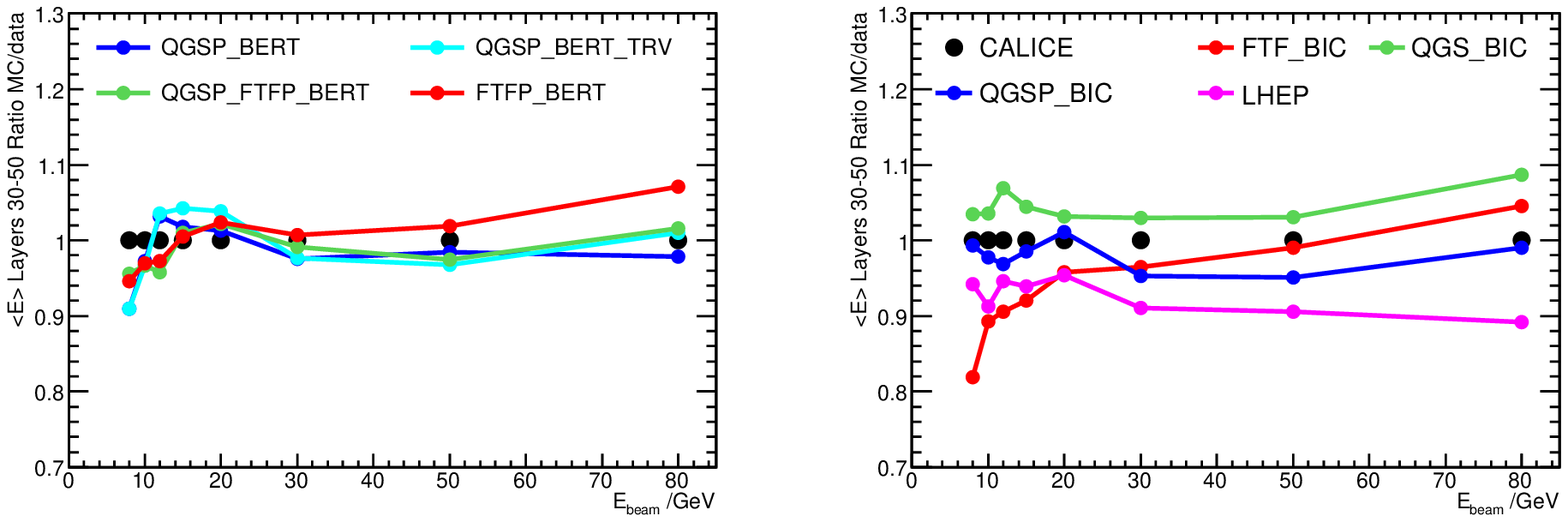}
\caption{\label{fig:LayerRatios}\em Ratio of simulation to data for 
three different regions of the longitudinal energy profile:  (top pair of plots) layers 1-3, 
dominated by nuclear breakup; (centre pair) layers 5-20, dominated by electromagnetic showers;
and (bottom pair) layers 30-50, dominated by penetrating hadrons.}
\end{figure}

In order to quantify these observations, and extend them to all energies and models,
we show in figure~\ref{fig:LayerRatios} ratios of simulation to data as a function of 
energy for three regions of the 
longitudinal profile: layers 1-3 (where the contribution from nuclear breakup 
dominates), layers 5-20 (dominated by the showering of electrons and photons) and
layers 30-50 (the tail, dominated by penetrating hadrons).  This makes it clear that
the greatest differences between the models occur in the first few layers, where
discrepancies up to $\sim20$~\% are observed.  The  {\tt FTFP\_BERT} and  {\tt FTF\_BIC} 
models lie consistently above the data, while   {\tt LHEP},  {\tt QGSP\_BIC}
and  {\tt QGS\_BIC} lie consistently below.  The other physics lists
make a transition between the two regimes, as their model content changes with energy. 
In the 
electromagnetic-dominated region, the most obvious outlier is {\tt LHEP}; the other models 
all agree with data  within $\sim15$\%, with
the {\tt FTFP\_BERT} and {\tt FTF\_BIC} models giving the best description.
In the tails, most models lie within $\sim$10\% of data; {\tt LHEP} is consistently low, as is
{\tt FTF\_BIC} at lower energies.  

On balance, it appears that the {\tt FTFP\_BERT} physics list, 
while not perfect, gives the best overall description of the longitudinal 
development of these showers.  We emphasise, however, that this remark 
refers only to the early part of the shower which is developed in the ECAL; we are not 
sensitive to the later parts of the shower.    

\subsection{Physics lists under development}
\label{sect:Chips}

There are several physics lists in {\tt GEANT} 4.9.3 which use the {\tt CHIPS} 
model~\cite{ref:CHIPS} alone or in conjunction with other models.   This model has 
undergone substantial development recently, and is now capable of modelling 
all parts of the hadronic interaction process.  We made a number of studies 
of these physics lists using a $\beta$-test version 4.9.3.b01 of {\tt GEANT}.
The results were very encouraging, and one of the physics lists tested, 
{\tt QGSC\_CHIPS}, gave a better description of our data than any of the 
physics lists discussed here.   However, the {\tt CHIPS} model is still being tuned, 
and in 
the released version of {\tt GEANT} 4.9.3, the {\tt CHIPS}-based models are 
less successful when confronted with our data.  
It would be premature to show results and draw conclusions while development is ongoing, 
but the {\tt CHIPS}-based 
models seem to be an interesting and promising new avenue, and it is clear that 
our data have the power to discriminate between tunings of such models.


\section{Summary}
\label{sect:Summ}
We have studied showers induced by charged pions in the CALICE silicon-tungsten ECAL. The calorimeter
has high transverse and longitudinal granularity, which  allows us to 
study the early part of the shower development in unprecedented detail.
Using pions in the energy range 8--80~GeV, 
we have compared the data with {\tt GEANT4} simulations 
using eight different physics lists.   Several observables were examined -- the total
energy deposited in the ECAL, the distribution of interaction points, 
the transverse shower energy profile, and the longitudinal profile with 
respect to the interaction point.    The most sensitive observables seem 
to be the shower profiles.  The data tend to show a greater shower width 
than the simulation for most energies and  physics lists.  The closest
description of the transverse profiles is provided by the
{\tt FTFP\_BERT} and {\tt FTF\_BIC} physics lists.   

The longitudinal profile seems the most interesting observable, because the three main
components of the shower induced by the primary interaction (energetic hadrons, 
photons and low energy nuclear fragments) can be, to some extent, distinguished through 
their different rates of shower development. An ECAL using tungsten is particularly
useful in this regard, because of its small ratio of $X_0/\lambda_{\mathrm{int.}}$, which 
amplifies the differences in shower development between various components of the shower.
It would be na\"{i}ve to expect any of
the physics lists to give a perfect description of the data, but it seems clear from 
our study that the {\tt LHEP} physics list has serious deficiencies, and that, 
in the framework of the current version 4.9.3 of {\tt GEANT4},  
{\tt FTFP\_BERT} list is the most successful.  It is to be hoped that these observations
will provide useful guidance to the {\tt GEANT4} developers.   

\section*{Acknowledgements}

We would like to thank the technicians and the engineers who
contributed to the design and construction of the prototypes.
We also
gratefully acknowledge the DESY and CERN managements for their support and
hospitality, and their accelerator staff for the reliable and efficient
beam operation. 
We would like to thank the HEP group of the University of
Tsukuba for the loan of drift chambers for the DESY test-beam.
The authors would like to thank the RIMST (Zelenograd) group for their
help and sensors manufacturing.
This work was supported by the 
Bundesministerium f\"{u}r Bildung und Forschung, Germany;
by the DFG cluster of excellence ``Origin and Structure of the Universe'';
by the Helmholtz-Nachwuchsgruppen grant VH-NG-206;
by the BMBF, grant numbers 05HS6VH1 and 05HS6GU1;
by the Alexander von Humboldt Foundation (Research Award IV, RUS1066839 GSA);
by joint Helmholtz Foundation and RFBR grant HRJRG-002, Russian Agency for Atomic Energy, ISTC grant 3090;
by Russian Grants  SS-1329.2008.2 and RFBR0402/17307a
and by the Russian Ministry of Education and Science;
by CICYT,Spain;
by CRI(MST) of MOST/KOSEF in Korea;
by the US Department of Energy and the US National Science
Foundation;
by the Ministry of Education, Youth and Sports of the Czech Republic
under the
projects AV0 Z3407391, AV0 Z10100502, LC527, LA09042
and by the Grant Agency of the Czech Republic
under the project 202/05/0653; 
and by the Science and Technology Facilities Council, UK.


\end{document}